\documentclass[fleqn,usenatbib]{mnras}

\usepackage{newtxtext,newtxmath}

\usepackage[T1]{fontenc}

\DeclareRobustCommand{\VAN}[3]{#2}
\let\VANthebibliography\thebibliography
\def\thebibliography{\DeclareRobustCommand{\VAN}[3]{##3}\VANthebibliography}

\usepackage{graphicx}	
\usepackage{amsmath}	
\usepackage{ulem}

\usepackage{soul}

\usepackage{listings}

\newcommand{\comment}[1]{}

\title[Orbital Structure Evolution in Self-Consistent N-body Simulations]{Orbital Structure Evolution in Self-Consistent N-body Simulations}

\author[D. Valencia-Enr\'{i}quez, I. Puerari, L. Chaves-Velasquez]{
Diego Valencia-Enr\'{i}quez$^{1}$,\thanks{E-mail: diegovalencia5@gmail.com (COL)}
Iv\^anio Puerari$^{2}$,
and Leonardo Chaves-Velasquez$^{3,4,5}$
\\
$^{1}$Universidad Mariana, Calle 18 No. 34 - 104, 52001	Pasto, Colombia\\
$^{2}$Instituto Nacional de Astrof\'{i}sica, \'{O}ptica y Electr\'{o}nica,  Calle Luis Enrique Erro 1, Santa Mar\'{i}a Tonantzintla, 72840 Puebla, Mexico\\
$^{3}$Instituto de Radioastronom\'{i}a y Astrof\'{i}sica, Universidad Nacional Aut\'{o}noma de M\'{e}xico, Apdo. Postal 3-72, Morelia, 58089 Michoac\'an, Mexico\\
$^{4}$Astronomical Observatory, Universidad de Nariño, Sede VIIS, Avenida Panamericana, Pasto, Nariño, Colombia\\
$^{5}$Departamento de Física de la Universidad de Nariño, Torobajo Calle 18 Carrera 50, Pasto, Nariño, Colombia
}

\date{Accepted XXX. Received YYY; in original form ZZZ}

\pubyear{2015}

\begin{document}
\label{firstpage}
\pagerange{\pageref{firstpage}--\pageref{lastpage}}
\maketitle

\begin{abstract}
The bar structure in disk galaxies models is formed by different families of orbits; however, it is not clear how these families of orbits support the bar throughout its secular evolution. Here, we analyze the orbital structure on three stellar disk N-body models embedded in a live dark matter halo. During the evolution of the models, disks naturally form a bar that buckles out of the galactic plane at different ages of the galaxy evolution generating boxy, X, peanut, and/or elongated shapes. To understand how the orbit families hold the bar structure, we evaluate the orbital evolution using the frequency analysis on phase space coordinates for all disk particles at different time intervals. We analyze the density maps morphology of the 2:1 family as the bar potential evolves. We showed that the families of orbits providing bar support exhibit variations during different stages of its evolutionary process, specifically prior to and subsequent to the buckling phase, likewise in the secular evolution of the bar. The disk-dominated model develops an internal boxy structure after the first Gyr. Afterwards, the outer part of the disk evolves into a peanut-shape, which lasts till the end of the simulation. The intermediary model develops the boxy structure only after 2 Gyr of evolution. The peanut shape appears 2 Gyr later and evolves slowly. The halo-dominated model develops the boxy structure much later, around 3 Gyr, and the peanut morphology is just incipient at the end of the simulation.
\end{abstract}




\section{Introduction}

In the context of the cosmological evolution of galaxies, baryonic matter disks assemble at the center of dark matter halos. In the absence of large perturbations, these disks undergo secular evolution, giving rise to several exciting structures, such as grand design spiral arms, flocculent spiral patterns, bars, lenses, and rings. Bars significantly impact the dynamics of the disks as they strongly alter the axisymmetric potential of the original disks \citep{2013seg..book..305A}.

In the local Universe, observers have noted that two-thirds of disk galaxies display bars \citep{2015ApJS..217...32B}. Optical surveys reveal that approximately 30 percent of these galactic bars are strong. If we also consider weak bars visible in the Fourier decomposition of the light distribution, the fraction increases to 50 percent or more \citep{2007ApJ...659.1176M}. This fraction even rises to $60 - 70$ percent when observing in the near-infrared \citep{2000AJ....119..536E,2009A&A...495..491A}. Furthermore, numerical simulations have demonstrated that bars grow in the vertical direction by a buckling instability, which then settles down into a boxy or peanut-like (B/P) shape when viewed edge-on \citep[e.g.][]{1991Natur.352..411R,1994ApJ...425..551M,2004ApJ...604L..93D,2004ApJ...613L..29M,2006ApJ...645..209D,2006ApJ...637..214M,2013ApJ...764..123S}. These structures are certainly found in observations \citep{2000A&AS..145..405L,2017MNRAS.468.2058E}. Yet barred galaxies show a pronounced B/P-shape (e.g., SDSS image of NGC 128), other galaxies do not display any boxy or peanut-like shape or X-shaped structure in their isophotes. These face-on bars only demonstrate elongated elliptical isophotes, e.g., IC 5240 \citep{2017A&A...598A..10L}. In this regard, \cite{2017ApJ...845...87L} conducted a statistical study of bar galaxies based on the Carnegie-Irvine Galaxy Survey (CGS), classifying such bars as unbuckled bars, B/P-shaped bars, and bars with barlenses, which are buckled bars. The study found that unbuckled bars do not exhibit any distinct features, except for elongated elliptical isophotes and are predominantly found in later-type galaxies, while buckled bars develop more frequently in early-type galaxies with a massive disk \citep{2017ApJ...845...87L,2017MNRAS.468.2058E}.

Researchers have studied bar formation in disk galaxies for several decades. Essentially, a bar-like structure is triggered by two processes: the first one stems from internal causes, such as dynamical instabilities within individual galaxies, and the second one is influenced by external (e.g., tidal) influences. \cite{1996LNP...474....7L} and \cite{2003AstL...29..447P} reviewed different mechanisms for bar formation that are all connected by the exchange of angular momentum. In numerical simulations, the evolution of bars in N-body models that are unstable during their formation depends on the distribution of stellar orbits. The different components of a galaxy (halo, bulge, and disk) interact with each other and cause the angular momentum exchange between disk resonances and their other components (\cite{2002ApJ...569L..83A} and references therein). 

N-body numerical simulations show that the vertical buckling instability influences the secular growth of stellar bars \citep{1981A&A....96..164C,1991Natur.352..411R,2006ApJ...637..214M}. \cite{2006ApJ...637..214M} showed the vertical structure of the bar buckles twice during the simulation. The first buckling appears in the central area of the bar, and the second buckling comes into view on the peripheral part of the bar. In that respect, \cite{2006ApJ...637..214M} analyzed the self-consistent evolution of the stellar bar to understand the three-dimensional structure by studying periodic orbits in arbitrary gravitational potentials. They chose periodic orbits, which characterize the overall orbital structure of the bar phase space, and showed that the general shape evolves from a peanut/boxy to a vertically asymmetric shape and then to a peanut/X-shape.

Those studies show that the general morphology of a barred galaxy changes during its evolution; therefore, the bar potential changes as well. So, different orbital families, which are the backbone of a barred structure, change their importance during the evolution of the bar. \cite{2018MNRAS.481.4058S, 2019MNRAS.485.1900S} demonstrate a relationship between the face-on and edge-on morphology of bars and the parameters of the underlying galaxy. Additionally, in their study, \cite{2021MNRAS.502.4689S} showed boxy/peanut bars are supported by different families of orbits. In the early work by \cite{1980A&A....92...33C}, they showcased the basic families of periodic orbits ($x_1$, $x_2$, and $x_3$) in different isochrone bar model potentials. They showed that in weak bars, inside the inner ILR and outside the outer ILR, the orbits follow the periodic $x_1$ family, and between the ILRs, most orbits follow the $x_2$ family. However, if the bar is sufficiently strong, the $x_2$ family of orbits does not exist. Another study by \cite{1983A&A...127..349A} investigated the orbits in prolate heterogeneous bar models and classified them into A (confined to small radii, anti-aligned), B (elongated orbits aligned with the bar), and R (retrograde orbits). They found that in the case of a massive bar, most quasi-periodic orbits belong around either the B or R family. Subsequent studies by \cite{2002MNRAS.333..847S,2002MNRAS.333..861S}, \cite{2002MNRAS.337..578P}, and \cite{2019MNRAS.490.2740P} revealed various bifurcations of the $x_1$ orbit families in 2D and 3D (known as the $x_1$-tree) that support the bar. They utilized Ferrers bar potentials with different pattern speeds to investigate these bifurcations. Similar studies have used frozen potentials of N-body simulations, as demonstrated by \cite{2013seg..book..305A} and \cite{2018A&A...612A.114P}. Other studies, such as those by \cite{1994LNP...430..264W}, \cite{1996A&A...309..381K}, \cite{1997ApJ...483..731P}, \cite{1999CeMDA..73..149W}, \cite{2005CeMDA..91..173M}, \cite{2009MNRAS.394.1605H}, \cite{2010MNRAS.408...22P}, \cite{2013MNRAS.436.1201C}, \cite{2014MNRAS.445.3546P}, \cite{2015MNRAS.448.3081T}, and \cite{2017ApJ...850..145C}, have demonstrated that sticky-chaotic orbits are capable of supporting galactic bars. Additionally, \cite{2019MNRAS.490.2740P} considered distinct types of periodic orbits, which do not belong to $x_1$ family, but these could probably be suitable foundation building blocks for bars.

\begin{figure*}
\includegraphics[scale=0.8]{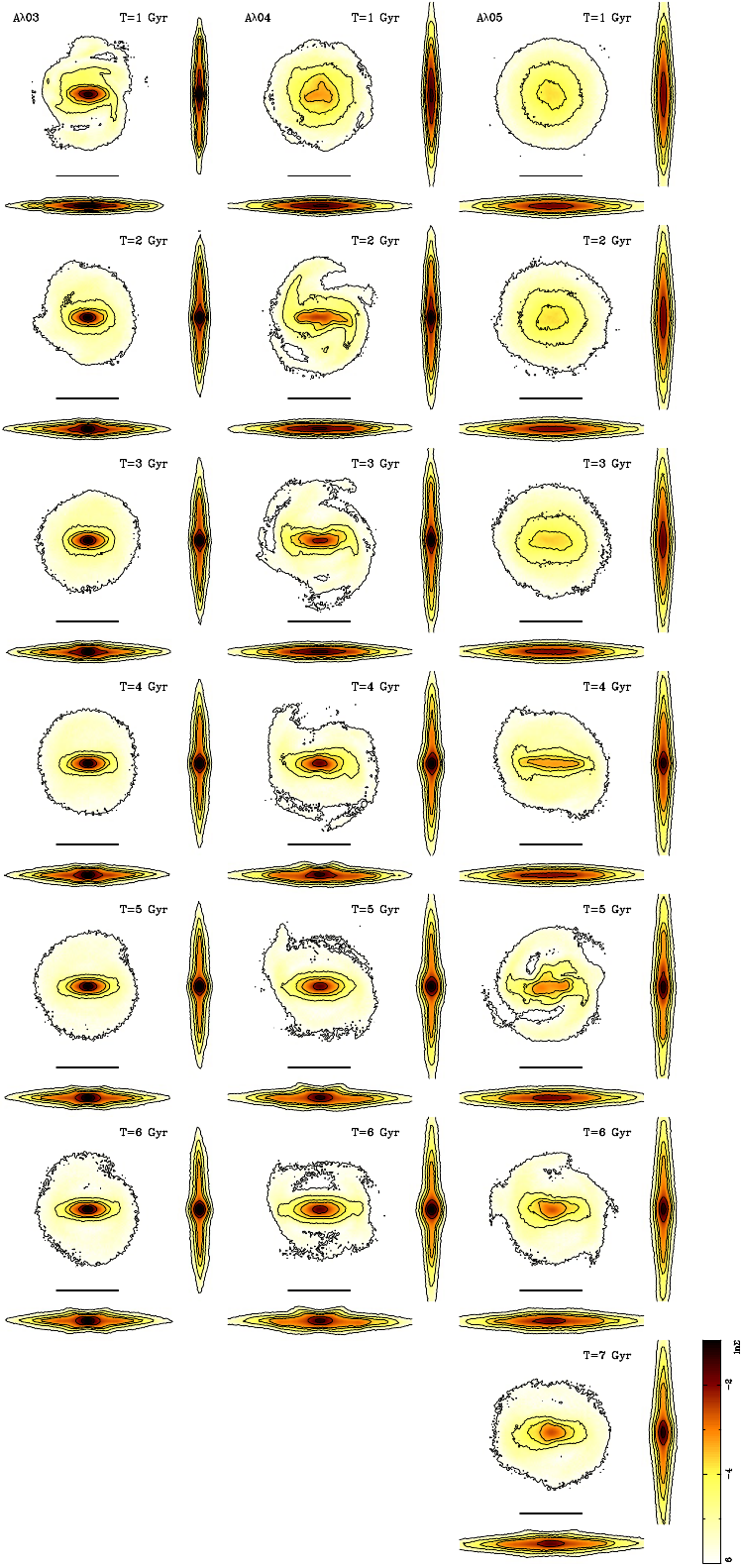}
	\caption{Surface logarithm density maps for our models and their evolution. The left, middle, and right panels show the models $A\lambda03$, $A\lambda04$, and $A\lambda05$, respectively. Time increases downward. The color scale at the bottom right of the figure corresponds to the logarithm of the surface density ($\Sigma$ in $1.0\times10^{10}$ M$_{\odot}$/kpc$^2$). Each panel shows the face-on, edge-on, and end-on views, with the horizontal lines below the face-on view indicating a distance of 10 kpc. From the figure, we can observe the bar growth and the formation of buckling. The buckling appears at different times in the models; around 2 Gyr for $A\lambda03$ and 4 Gyr for $A\lambda04$. For $A\lambda05$, the buckling is incipient at the end of the simulation (7 Gyr).}
	\label{fig:models}
\end{figure*}

The computed orbital frequency of the disk particles determines the orbital structure in barred galaxies. According to \cite{2007MNRAS.381..757V}, not only 2:1 resonance orbits but also orbits resonances close to 3:1 and 5:2 support the bar with regular and chaotic orbits. Furthermore, \cite{2009MNRAS.394.1605H} presented the distribution of frequencies for the most significant populations in various N-body simulations with frozen potentials. They demonstrated that B/P (boxy-peanut) bulges are composed of a superposition of several peanut shapes produced by different orbits, and peanut-shaped bulge is primarily assembled of families of pretzel-like orbits. Additionally, \cite{2016ApJ...818..141V}, and \cite{2017MNRAS.470.1526A} showed that the fraction of orbits associated with different families may somewhat depend on the details of the bar potential or the parameters of the underlying galaxy \citep{2020ApJ...895...12P}. Conducting a frequency analysis directly from the N-body model, \cite{2016ApJ...830..108G} found that stellar orbits have boxy shapes with significant elongation degree, exhibiting banana or infinity symbol shapes when viewed edge-on. Finally, \cite{2019A&A...629A..52L} suggests that the initial distortion of the bar is likely caused by the resonant trapping of $x_1$ orbits, which subsequently transition into banana-like orbits. In general, all these works highlight the existence of different orbit families that support the bar, but remains unclear which specific families of orbits support the bar as it evolves.

\begin{figure}
 \includegraphics[scale=0.35,angle=-90]{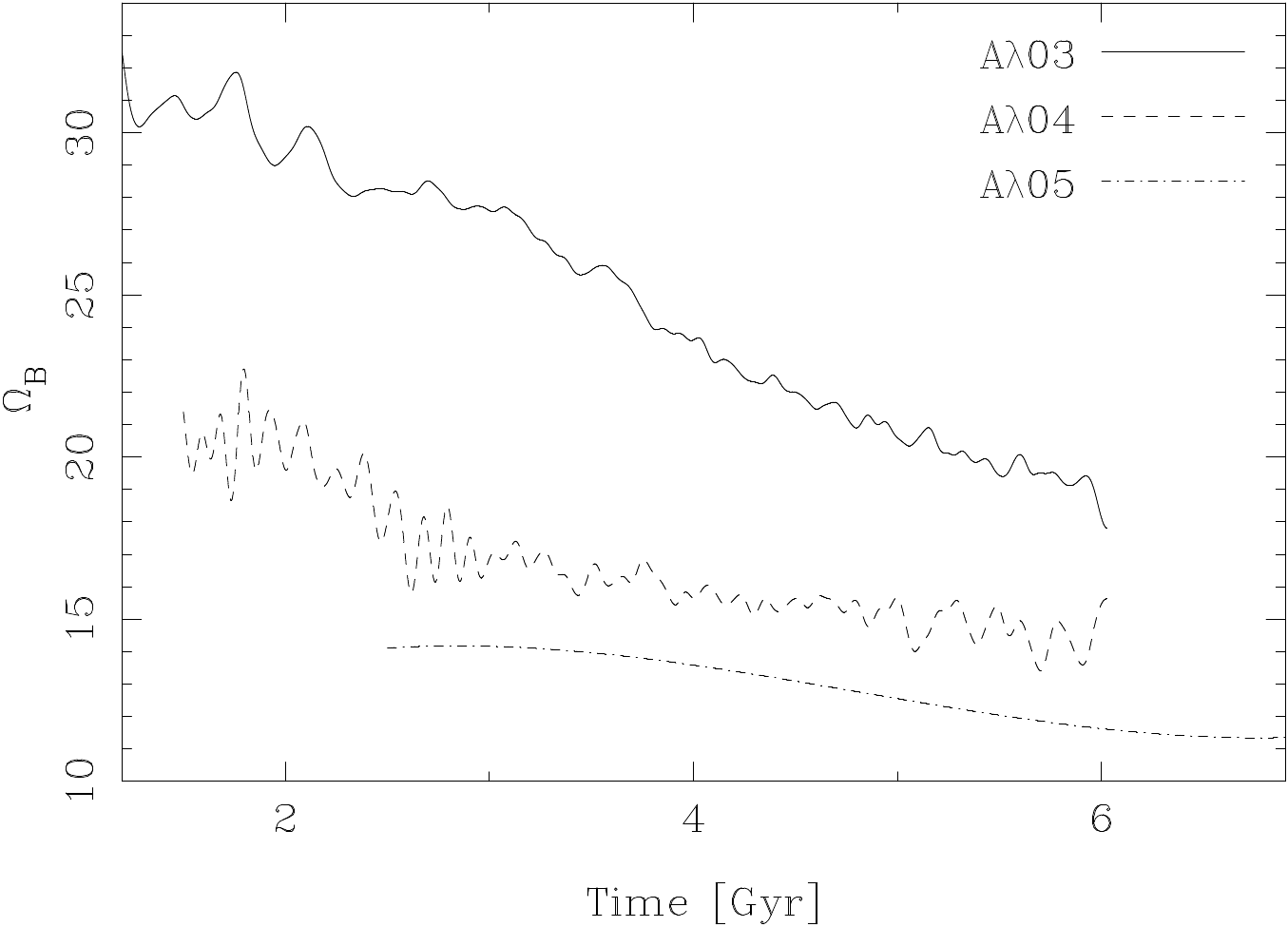}	
 \caption{Slowdown of the bar pattern speed for $A\lambda03$ (full line), $A\lambda04$ (dashed line), and $A\lambda05$ (dot-dash line) models.}
\label{fig:omegas}
\end{figure}

In this study, we show the evolution of the orbital structure in three fully self-consistent N-body models taken from \cite{2019AJ....157..175V}. To assess this evolution, we measure a range of frequencies for all disk particles at different time intervals. We examine the distribution of orbital frequencies across the face-on, edge-on, and end-on perspectives from the barred disk galaxy models as the bar potential undergoes changes. Section 2 provides an overview of the N-body models and the frequency analysis methodology employed in this research. We present our findings on the evolution of the orbital structure in Section 3. Finally, in Sections 4 and 5, we discuss our results and draw conclusions based on our analysis.

\section{N-Body Models and Methodology}

This section includes a description of galaxy models and the methodology to study the evolution of orbital structure through frequency analysis.

\subsection{Disk Galaxy Models}

We use the N-body simulations described in \citet{2019AJ....157..175V} (hereafter VE19) to analyze the orbital frequencies. VE19 presented two sets of simulations: four isolated models and thirteen perturbed ones. In this study, we focus on conducting a frequency analysis for three isolated models. The models begin as an axisymmetric disk, which, due to internal dynamical instabilities \citep[and references therein]{2002ApJ...569L..83A,2003MNRAS.341.1179A}, develop strong bars that persist until the end of the simulations. The isolated models employed an exponential stellar disk $\Sigma(r)=\Sigma_0 e^{r/r_d}$ where $\Sigma_0=M_d/(2\pi r_d)$, and an NFW dark matter halo \citep{1996ApJ...462..563N,1997ApJ...490..493N}. The vertical mass disk distribution is composed of an isothermal sheet with a radially constant vertical scale length $z_0$. Therefore, the three-dimensional stellar density in the disk is $\rho_d(r,z)=\Sigma(r)/(2z_0)sech^2(z/(2z_0))$.

The models were built as an equilibrium N-body realization \citep{1999MNRAS.307..162S} of $7\times10^{6}$ particles ($2\times10^{6}$ for the disk and $5\times10^{6}$ for the halo). The halo has a mass of $5.11\times10^{11}$ $M_{\odot}$ and a concentration of 8.0. The disk has a mass of $2.55\times10^{10}$ $M_{\odot}$ and radial scale-length changes for the different models. For the $A\lambda03$, $A\lambda04$, and $A\lambda05$ models discussed here, the radial scales are $r_d=1.99$, $r_d=2.80$, and $r_d=3.58$ kpc, respectively, and the models have $z_0=0.2r_d$. We performed collisionless N-body simulations with the Gadget-2 code \citep{2001NewA....6...79S,2005MNRAS.364.1105S}.

Models described in VE19 were chosen to represent galaxies where the disk dominates (the Critical Spin parameter $\lambda_c$ is larger than the Spin parameter of the disk $\lambda_d$) and models where the halo dominates ($\lambda_c < \lambda_d$). VE19 showed that the isolated model keeps that configuration during the whole evolution, and the disturbed models change such parameters at the interaction. Therefore, models form a bar if $\lambda_c > \lambda_d$ for both isolated and interaction models. Besides, disk-dominate models develop a bar relatively quickly, and halo-dominate ones do not form that structure. The disk-dominate models evolve strong bars due to the exchange of angular momentum between the components of the galaxy (e.i. halo and disk) and between the inner and outer region of the disk \citep{2013MNRAS.429.1949A}. This process is more efficient if the Critical Spin parameter is relatively much larger than the Spin parameter of the disk.

In this article, we recalculate the simulations of the isolated models $A\lambda03$, $A\lambda04$, and $A\lambda05$ from VE19, now with a higher time resolution. This is very important to have a better determination of the orbital frequencies (see Section \ref{sec:ose}).

\begin{figure*}
    \includegraphics[scale=0.8]{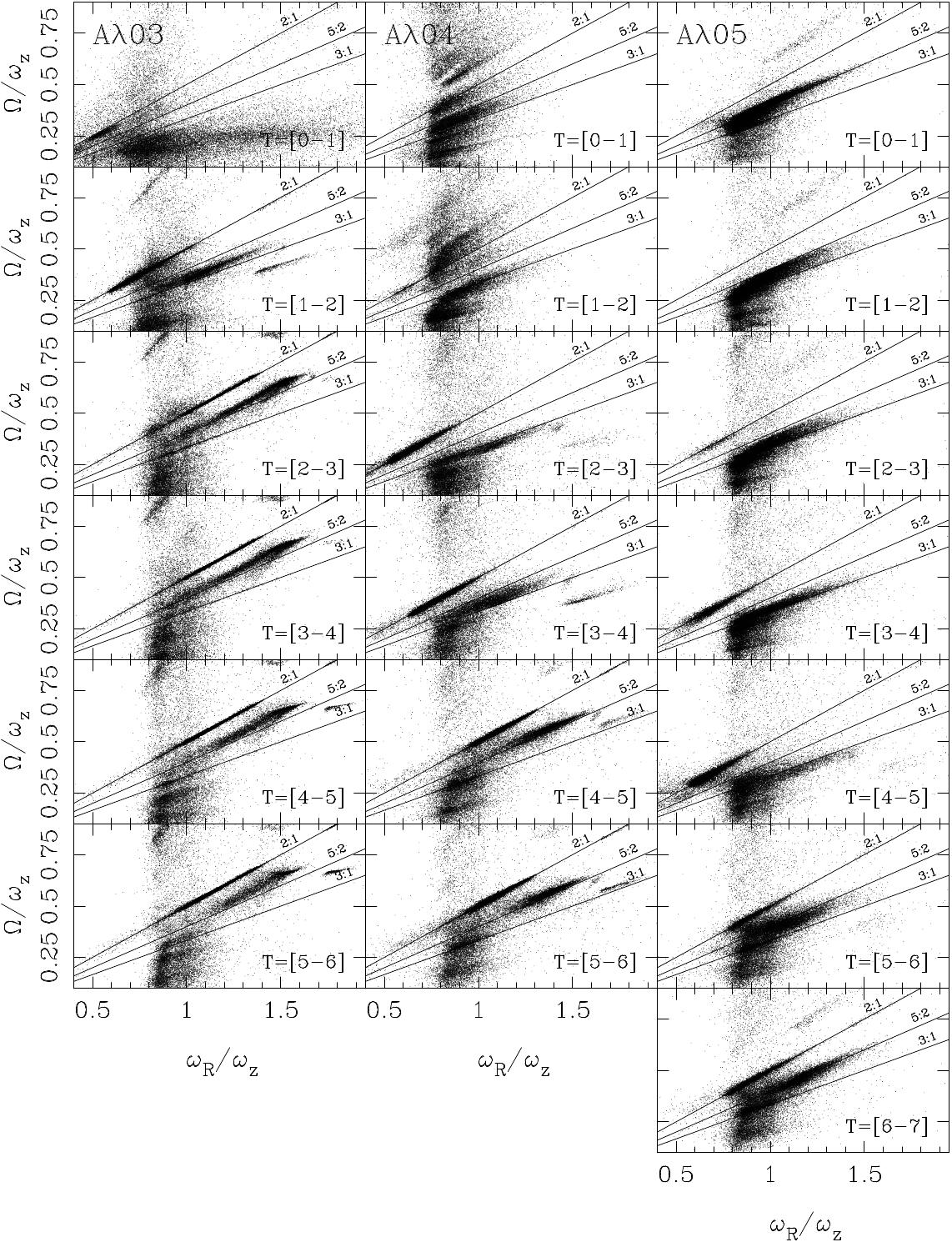}	
	\caption{The plots show the ratio frequency maps for all disk particles where the time intervals in which the frequencies were calculated increase from top to bottom. The left, middle, and right columns represent $A\lambda03$, $A\lambda04$, and $A\lambda05$ models, respectively. The straight lines in the plots show the location of the resonances 2:1, 5:2, and 3:1. We note for $A\lambda03$ that the 2:1 resonance is quickly populated, and the particles move to larger $\omega_R/\omega_z$ on this resonance as they evolve. The same resonance is populated later on in the $A\lambda04$ model and at much later times for $A\lambda05$ model. As noted in Section \ref{sec:ofa}, the value of $\Omega$ may have larger errors than the other frequencies.}
	\label{fig:all_MAP_ok_B}
\end{figure*}

\subsection{Frequency analysis method}

The orbits of stars in a galaxy describe oscillations in three dimensions. In Cartesian coordinates ($x$, $y$ and $z$), the $\omega_x$, $\omega_y$, and $\omega_z$ frequencies represent those oscillations. Likewise, in cylindrical coordinates, the $\Omega$ and $\omega_R$ frequencies represent the azimuthal and radial oscillations; $\omega_z$ represents the vertical oscillation for both coordinate systems.

In order to analyze these frequencies in our models, we take the bar frame of reference where the bar is in a horizontal position for all snapshots (see Figure \ref{fig:models}). We calculate the frequencies of the orbits from each disk particle using the phase space coordinates as a function of discrete time ($x(t)$, $y(t)$, $z(t)$, $v_x(t)$, $v_y(t)$ and $v_z(t)$ for Cartesian coordinates, and the azimuthal position $\theta(t)$ and the radial velocity $v_r(t)$, for cylindrical coordinates). 

We use an algorithm similar to the numerical analysis of fundamental frequencies, developed by \citet{1990Icar...88..266L}. To apply the frequency analysis, we find the peaks of amplitude from the Fourier Transform of a time series of a given coordinate, e.g., $x(t)$. The continuous Fourier Transform coefficients, $H_x(w)$, have the following form
\begin{equation}
H_x(\omega_x)=\left|\int_{-\infty}^{\infty}x(t)e^{i\omega_x t}dt\right|.
\end{equation}
Because $x(t)$ results from a simulation,  we have a finite number of sampled points where we have $N$ consecutive sampled values $x_k=x(t_k)$, equally spaced in the interval time $\Delta t$, so $t_k=k\Delta t$, and $k=0,1,2,...,N-1$. The integral approximates as the sum:
\begin{equation}
H_x(\omega_x)=\left| \dfrac{1}{N}\sum_{k=0}^{N-1}x_ke^{i\omega_xk\Delta t} \right|
\end{equation}

To perform the analysis, we first subtract the average value of $x(t)$ from the signal to eliminate the spurious amplitudes that can appear at the zero frequency. Next, we filter the signal by multiplying it with the Hanning window to reduce side lobes. Then, we apply the Fast Fourier Transform (FFT) for a real-valued sample using the \texttt{realft} subroutine of \citet{press1992numerical}. We search for a maximal value of the $H_x(w_k)$ then we save the points  $(w_k, H_x(w_k))$, the previous value $(w_{k-1},H_x(w_{k-1}))$, and the next value $(w_{k+1},H_x(w_{k+1}))$. With these three points, we fit a parabolic curve to get the vertex position. That value corresponds to the frequency $\omega_x$. We check our code reproducing the orbits showed in \cite{1998MNRAS.298....1C}, and we obtain the same frequencies; likewise, we test our code with a periodic function $x'(t_k)=H_x\sin(w_kt_k + \phi)$, and the calculated $w_k$ agrees very well with the input $w_k$.

\section{Orbital Structure Evolution}\label{sec:ose}

For a more precise frequency determination, we recalculated the simulations $A\lambda03$, $A\lambda04$, and $A\lambda05$ presented in VE19 with higher time resolution; instead of having 612 snapshots as in VE19, the simulations have 6120 steps to represent 6.0 Gyr in the simulations. Furthermore, we evolve the $A\lambda05$ model for one more Gyr because in this model, the bar formation is triggered much later compared to the other two models (Figure \ref{fig:models}). By having more data points for the same time interval, we ensure that the Nyquist critical frequency improves, and then the estimation of the frequencies of the orbits of the disk particles are better. In this way, for our simulations, we successfully calculated the frequencies using time intervals of 1 Gyr. In \cite{2016ApJ...830..108G}, the bar is tidally induced and presents a constant pattern speed during the time interval. The bars in our models are evolving, and their pattern speed changes slowly (see Figure \ref{fig:omegas}). This figure shows that the slowdown for the $A\lambda03$ model is approximately three km/s/kpc/Gyr, while for $A\lambda04$ and $A\lambda05$ is one km/s/kpc/Gyr approximately. So, the frequencies we calculated must be understood as the most predominant frequency of a given particle within a given time interval. In the following sections, we present our results for the three models, covering 6 (or 7 in the case of $A\lambda05$) time intervals.

\subsection{Frequency evolution of all disk orbits}\label{sec:ofa}

\begin{figure}
    \includegraphics[scale=0.3]{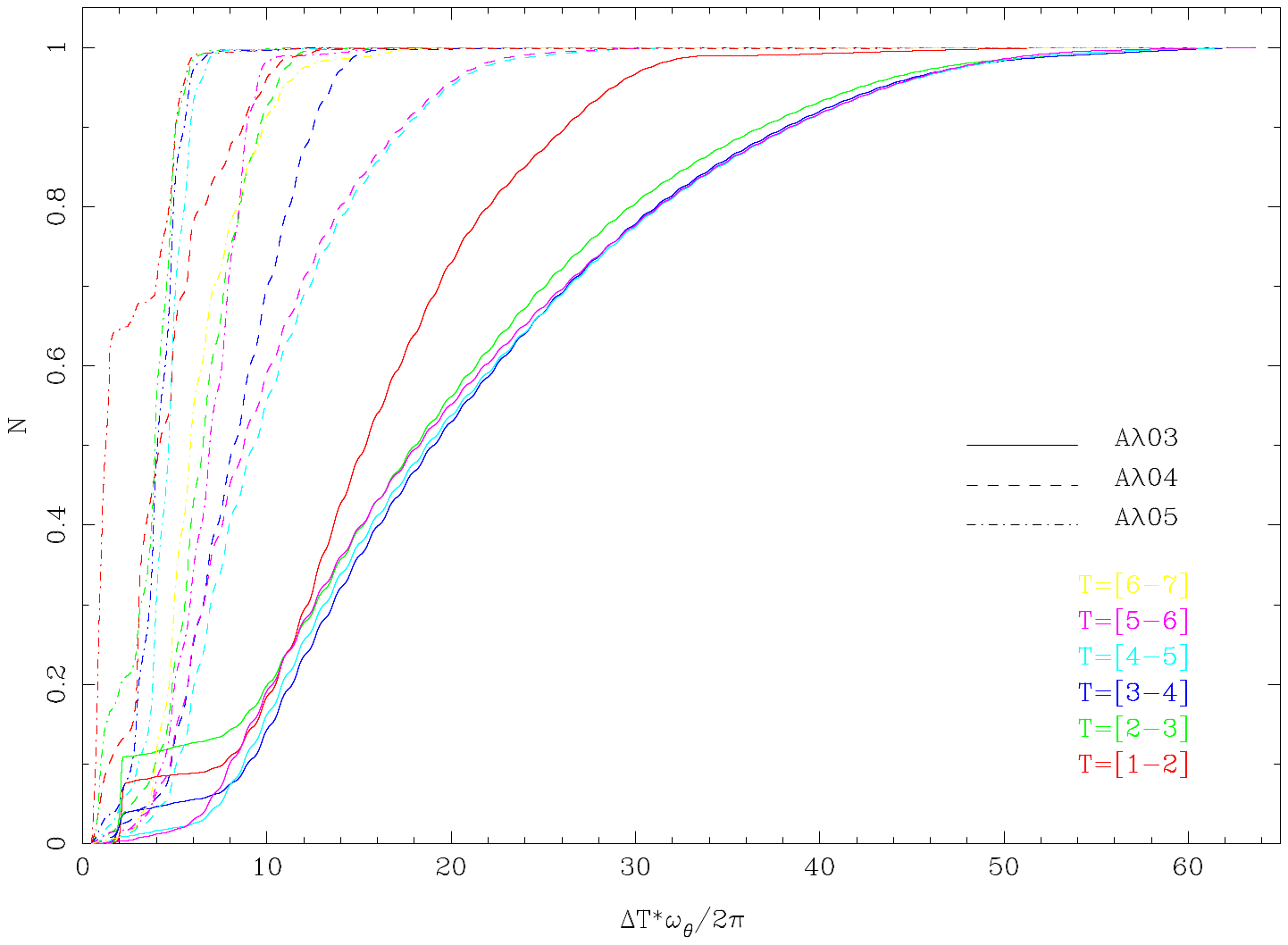}	
    \caption{Normalized cumulative number of orbits as a function of the number of periods of the particles for orbits within the frequency range $1.9<w_R/w_x<2.1$.}
	\label{fig:cumulative}
\end{figure}

We plot in Figure \ref{fig:all_MAP_ok_B}, the frequency ratio maps $\Omega/\omega_z$ versus $\omega_R/\omega_z$ for all disk particles of our models, and for all intervals of time. The frequency $\Omega$ here is just $\omega_\theta$, the frequency related to the cylindrical coordinate $\theta$. As discussed in \cite{2002ApJ...569L..83A},\cite{2007MNRAS.379.1155C}, and \cite{2016ApJ...830..108G}, the frequency $\omega_\theta$ is much more complex to calculate and in some cases, the direct Fourier Transform may result in erroneous values. In our case, we only plot this frequency in Figures \ref{fig:all_MAP_ok_B} and \ref{fig:cumulative}, and we continue our main analysis using the Cartesian coordinates. We observe that the bar structure initiates at different times for the various models. The 2:1, 5:2, and 3:1 resonances are distinguished in their population as the models evolve. In the case of the disk-dominated model $A\lambda03$, these resonances are swiftly populated. Conversely, for the other models, the population of these resonances occurs at later times.

\begin{figure*}
\includegraphics[scale=0.4]{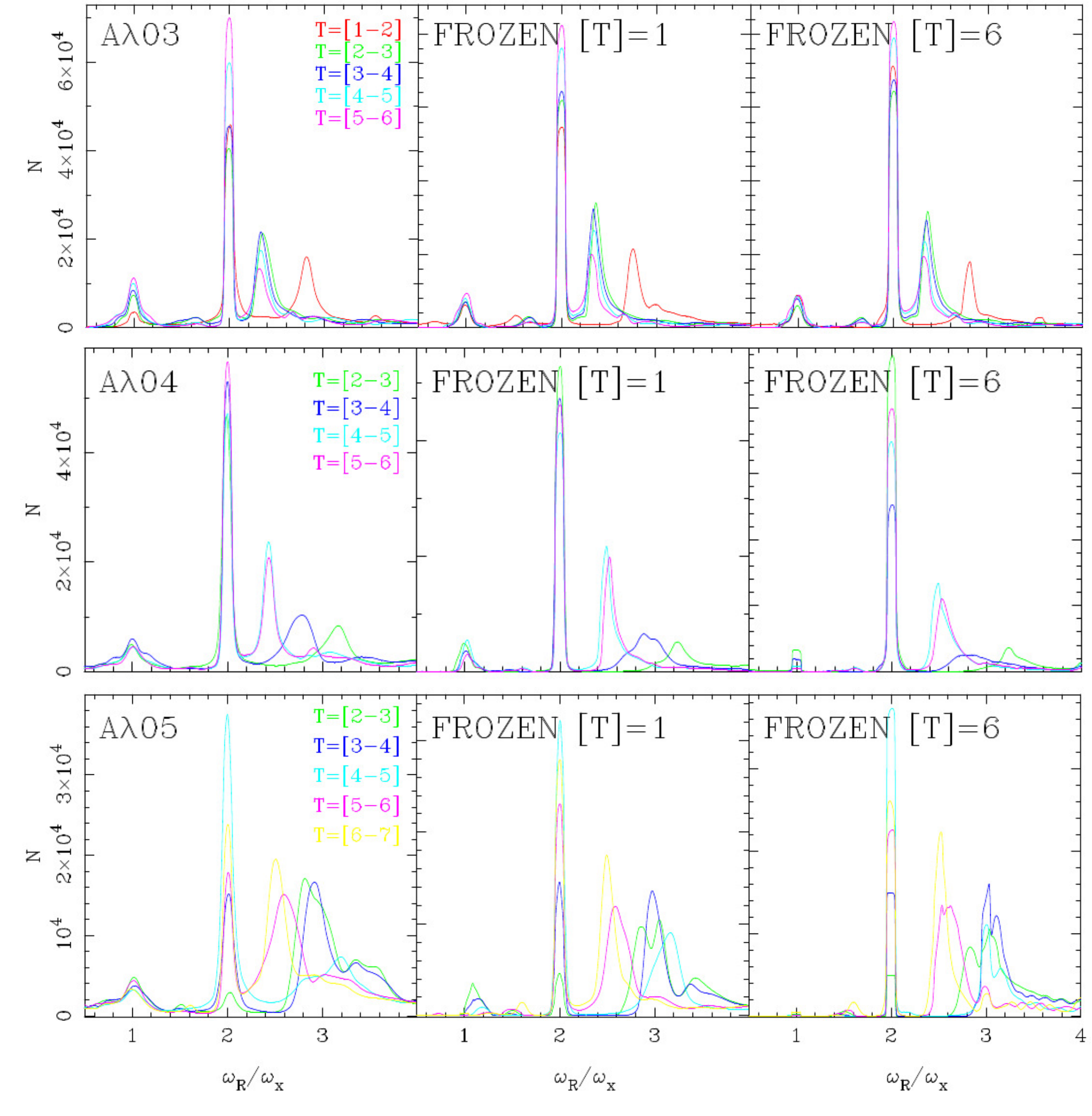}
    \caption{Histograms for the ratio $\omega_R/\omega_x$ for all disk particles for different time intervals. Top, middle, and bottom rows show the results for $A\lambda03$, $A\lambda04$, and $A\lambda05$ models, respectively, together with the results from the frozen potential calculations. We notice how the resonances are populated. The $A\lambda03$ shows a large number of particles on the 2:1 resonance and some around 5:2, and $A\lambda04$ follows a similar behaviour. Instead, $A\lambda05$ displays particles in a large range of $\omega_R/\omega_x$.}
	\label{fig:L03_frec_ry_B}
\end{figure*}

For $A\lambda03$ model (panels in the left column in Figure \ref{fig:all_MAP_ok_B}), in the first Gyr of the evolution, the resonance 2:1 begins to be populated with particles with low $\Omega/\omega_z$ and $\omega_R/\omega_z$: these particles have high $\omega_z$. Out of this resonance, the particles are distributed main at low $\Omega/\omega_z$ and at a large range of $\omega_R/\omega_z$. As the model evolves, particles begin to populate other resonances. For the time interval T=[1-2], both 2:1 and 3:1 are populated. From T=2 to the end of the evolution, a large number of particles are trapped in the 2:1 resonance and around the 5:2 one. Furthermore, note how these particles move on these resonances from low to high $\Omega/\omega_z$ and $\omega_R/\omega_z$. This movement is due to the decreasing of the vertical frequency $\omega_z$ of these particles as the buckling of the bar structure appears. Orbits not belonging to the main resonances do not suffer this effect. 

The panels in the middle column in Figure \ref{fig:all_MAP_ok_B} show the evolution for model $A\lambda04$. Here, at the early stages of the model evolution, particles are caught at some resonances (note the radial structures in T=[0-1] and T=[1-2]). At the time interval T=[2-3], the resonances 2:1 and 3:1 begin to be populated. As in the model $A\lambda03$, particles here also move from the 3:1 to the 5:2 resonances and suffers the same effect we have noticed: the particles move on these resonances from low to high $\Omega/\omega_z$ and $\omega_R/\omega_z$ as the buckling appears.

Finally, panels in the right column of Figure \ref{fig:all_MAP_ok_B} represent the evolution of the halo-dominated $A\lambda05$ model. In this case, numerous particles stay around the 5:2 and the 3:1 resonances for the first 3 Gyrs of the simulation. Only then the main 2:1 resonance begins to be populated. Moreover, the particles on these resonances also undergo the effect of moving to large values of $\Omega/\omega_z$ and $\omega_R/\omega_z$, but the effect is much less important. This is attributed to the very late appearance of the buckling phase in this model.

To understand the number of azimuthal periods used in the calculation of particles frequencies, we constructed an accumulative figure representing the number of orbits normalized to one as a function of the azimuthal rotations of the particle orbits (see Figure \ref{fig:cumulative}).
It is important to note that, in order to obtain an accurate frequency calculation using the Fourier Transform, the signal must contain at least one complete cycle within the analyzed time interval. This implies that the minimum number of periods required in the signal depends on the highest frequency present in the signal and the duration of the analyzed time interval. The highest frequency in the signal is related to the sampling frequency used in the data acquisition process. According to the Nyquist-Shannon sampling theorem, to avoid the phenomenon known as "aliasing" and accurately recover the frequencies present in the signal, the sampling frequency must be at least twice the highest frequency present in the signal \citep{oppenheim1997signals}.

In the case of models $A\lambda03$, $A\lambda04$, and $A\lambda05$, the frequency calculations are accurate because over 97\% of the particles show two or more rotations. Specifically, for model $A\lambda03$, more than 50\% of the particles exhibit over twenty cycles; for model $A\lambda04$, more than 50\% of the particles have approximately eight cycles or more; and for model $A\lambda05$, 50\% of the particles have five rotations (see Figure \ref{fig:cumulative}).

To better quantify the number of particles that are grabbed by the bar potential as the models evolve, we describe the evolution of the frequencies ratio $\omega_R/\omega_x$ as the bar growth. Figure \ref{fig:L03_frec_ry_B} shows the histograms for the $\omega_R/\omega_x$ quotient at different time intervals for our models and the Frozen potential experiments. The left column of this figure shows the results of the N-body models, while the center and right columns show the results for the frozen potential. The frozen potentials were calculated using a snapshot from the middle of each time interval. The orbits are then integrated during 1 Gyr (center column) and 6 Gyr (right column). The details of the Frozen potential calculations are described in Appendix \ref{appendix0}. 

For the disk-dominated model $A\lambda03$, during the T=[1-2] interval (marked by the red line), the model exhibits a significant peak at $\omega_R/\omega_x\approx2.0$ and a secondary peak at $\omega_R/\omega_x\approx2.8$. The latter peak corresponds to particles trapped around the 3:1 resonance. Notably, particles trapped around the 2:1 and 3:1 resonances already display a prominent bar structure in the model (refer to Figure \ref{fig:models} and Figure \ref{fig:maps}). As the model evolves, the peak around $\omega_R/\omega_x\approx2.0$ grows, and the 3:1 resonance particles transition into a 5:2 configuration (indicated by the peaks at $\omega_R/\omega_x\approx2.4$). Additionally, we observed a decrease in the significance of the peak around the 5:2 resonance, while the 2:1 resonance increasingly traps more particles.

The evolution of the $\omega_R/\omega_x$ ratio for the $A\lambda04$ model is shown in the middle row of Figure \ref{fig:L03_frec_ry_B}. It is only at the interval T=[2-3] when the resonance 2:1 begins to be populated (green line). For latter time intervals, the evolution is similar to the $A\lambda03$ model the peak at $\omega_R/\omega_x\approx2.0$ increases, while particles from the 3:1 resonance move to the 5:2 one.

The evolution for the $\omega_R/\omega_x$ ratio for the halo-dominated model $A\lambda05$ (bottom row of Figure \ref{fig:L03_frec_ry_B}) is quite different. For instance, the histogram shows a considerable amount of particles, which are around $\omega_R/\omega_x\approx3.0$, but these particles do not exhibit any barred structure (see Figure \ref{fig:models}). Then, the amount of these orbits decrease (see the cyan line) as the peak moves for lower $\omega_R/\omega_x$, up to the appearance of the oval distortion about the fourth Gyr (cyan line). In the next time interval (T=[5-6]), the bar seems to lost particles in the $\omega_R/\omega_x\approx2.0$ quotient; however, at the next time interval, this peak begins to be more significant again. On the other hand, the behavior depicted in the results of the frozen potentials (see second and third columns of Figure \ref{fig:L03_frec_ry_B}) is similar to the N-body models. The importance of the peaks appears in the same resonances $\omega_R/\omega_x$ that the N-body models for both time integrations (1 and 6 Gyr), showing that 1 Gyga-year is sufficient to get the frequencies in an N-body model. The slight change in the pattern speed rotation for an interval of 1 Gyr does not considerably affect the results when we measure the frequencies from an N-body model.

\subsection{Frequency evolution of Bar type orbits}\label{sec:febo}

\begin{figure}
	\includegraphics[scale=0.67,angle=0]{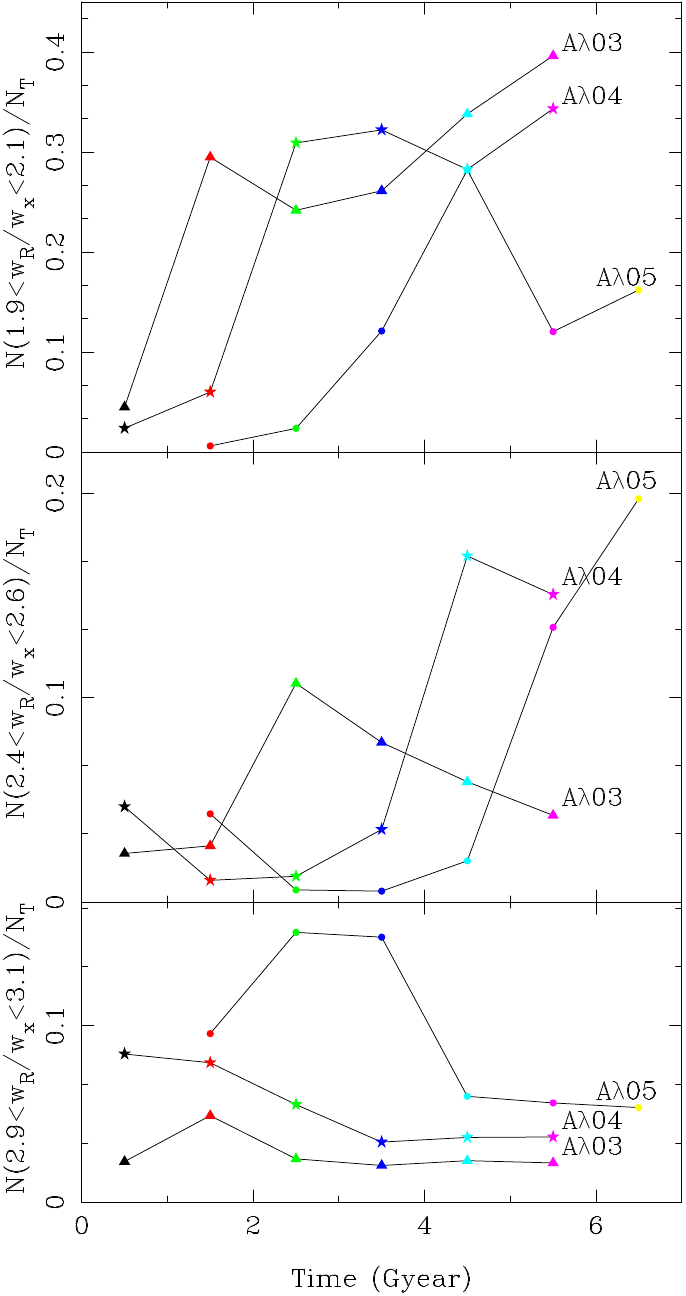}
	\caption{Disk mass fraction of particles with frequencies showing the ratio of $1.9<\omega_R/\omega_x<2.1$ (upper panel), $2.4<\omega_R/\omega_x<2.6$ (middle panel), and $2.9<\omega_R/\omega_x<3.1$ (bottom panel). The filled triangle, star, and circle symbols represent the $A\lambda03$, $A\lambda04$, and $A\lambda05$ models, respectively. The colors represent the time interval as in Figure \ref{fig:L03_frec_ry_B}. We can notice that the $A\lambda04$ model has the highest amount of B-type particles between $2-4$ Gyrs, which corresponds to the bar saturation phase in that model. At the same time, the bar formation is just starting for the $A\lambda05$ model.}
	\label{fig:Nwrwy}
\end{figure}

\cite{2015MNRAS.450L..66P} showed that orbits with an elongated shape along the bar major axis are identified by the frequency ratio $\omega_R/\omega_x\approx2$, which are related to orbits on the 2:1 resonance. Therefore, we choose the orbits in the interval $1.9<\omega_R/\omega_x<2.1$ and study their evolution. Figure \ref{fig:Nwrwy} shows the disk mass fraction of orbits between $1.9<\omega_R/\omega_x<2.1$ ratio, hereafter B-type orbits (top panel). Additionally, we show the disk mass fraction to orbtis between $2.4<\omega_R/\omega_x<2.6$ (middle panel), and $2.9<\omega_R/\omega_x<3.1$ (button panel). The filled triangle, star, and circle symbols represent the $A\lambda03$, $A\lambda04$, and $A\lambda05$ models, respectively. The color of the symbols depicts the time interval as in Figure \ref{fig:L03_frec_ry_B}. 

\begin{figure*}
	\includegraphics[scale=0.9]{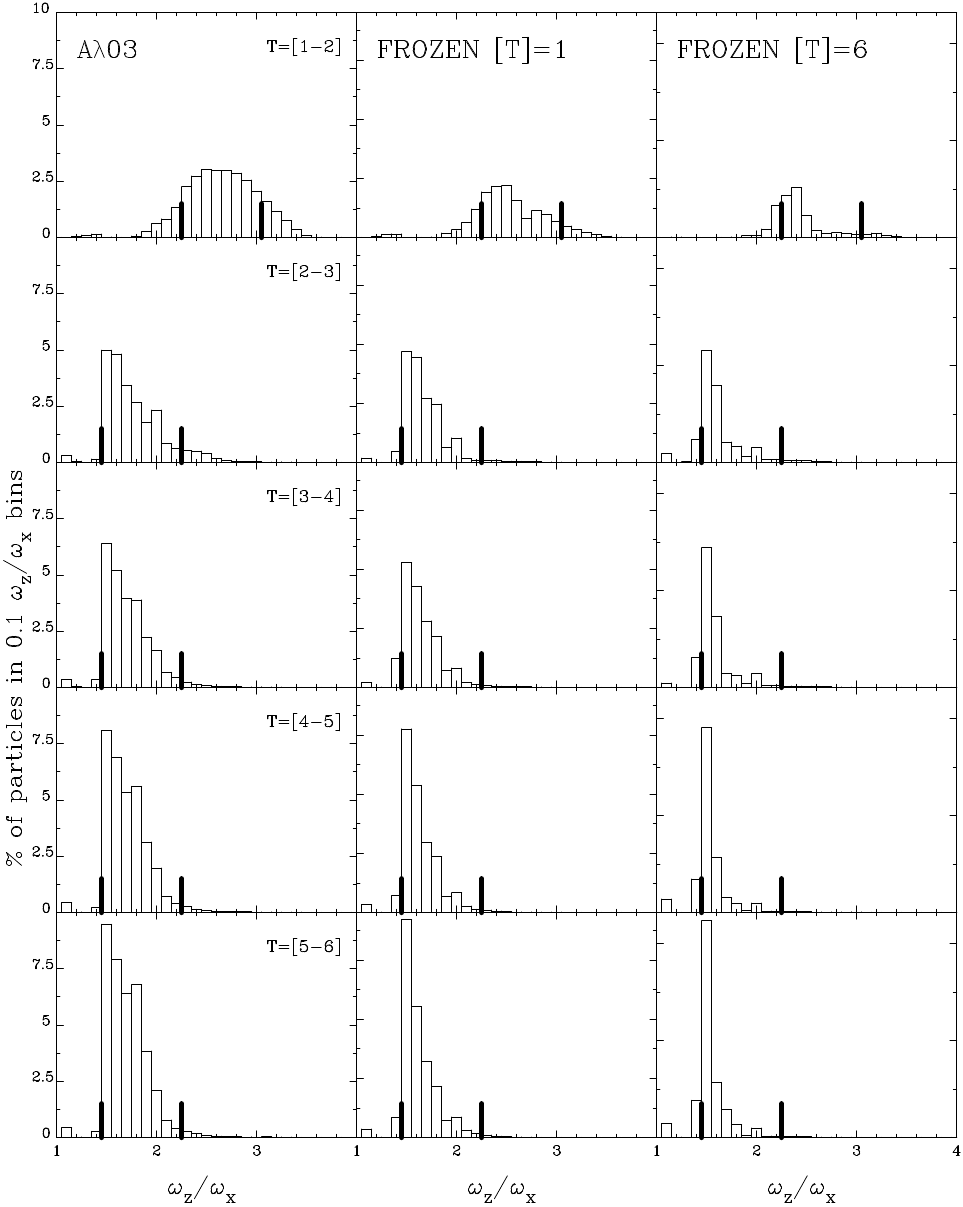}
	\caption{Histograms were generated to analyze the ratio $\omega_z/\omega_x$ for orbits within the range of $1.9<\omega_R/\omega_x<2.1$ (B-type orbits). The left column illustrates the evolutionary behavior of orbital frequencies in the $A\lambda03$ N-body model. The middle column depicts frequency analysis performed on the frozen potentials integrated throughout 1 Gyr. The right column showcases the frequency measures integrated over a time span of 6 Gyr. In this figure, time progresses downwards. The two prominent vertical lines in the histogram represent the range of $\omega_z/\omega_x$ used to construct the density maps (Figures \ref{fig:maps}, \ref{fig:L03_map}, \ref{fig:L04_map}, and \ref{fig:L05_map}).}
	\label{fig:frozenL03}
\end{figure*}

\begin{figure*}
	\includegraphics[scale=0.9]{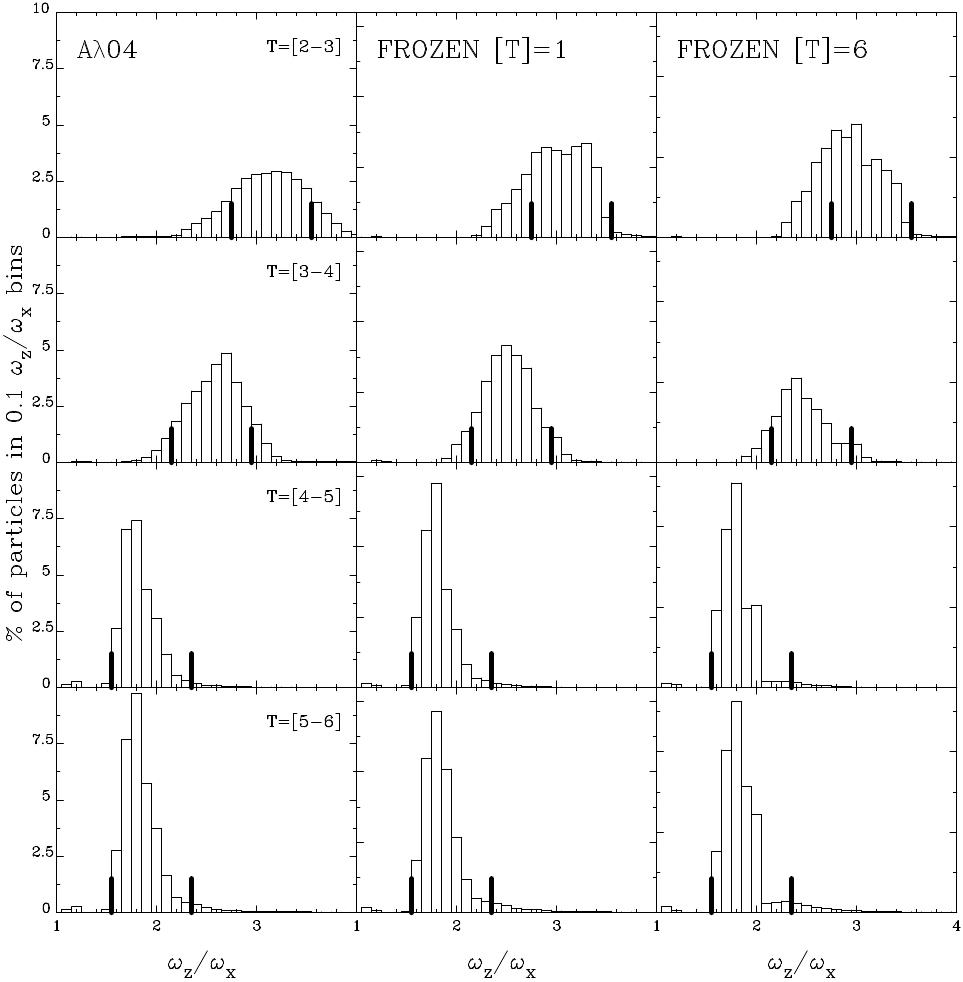}
	\caption{As in Figure \ref{fig:frozenL03}, but for model $A\lambda04$.}
	\label{fig:frozenL04}
\end{figure*}

\begin{figure*}
	\includegraphics[scale=0.9]{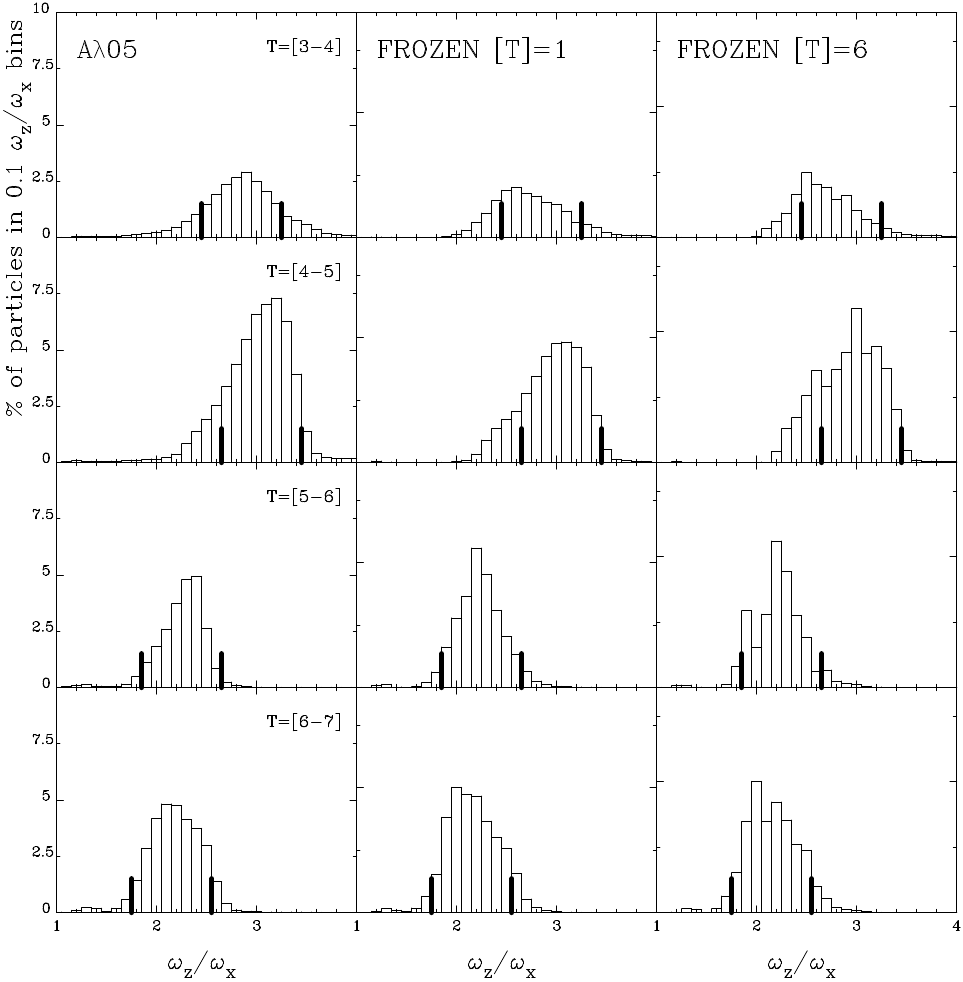}
	\caption{As in Figure \ref{fig:frozenL03}, but for model $A\lambda05$.}
	\label{fig:frozenL05}
\end{figure*}

The growth of B-type orbits in our models, depicted in the top panel of Figure \ref{fig:Nwrwy}, is different; it depends on the initial spin parameter of the disk $\lambda_d$. For the $A\lambda03$ model, these orbits reach a value around 30\% of the total disk mass fraction at the bar saturation phase\footnote{Bar saturation phase is the time when the amplitude of 1-D Fourier Transform coefficients for mode $m=2$ for disk particles is the maximum (see Figure 8 from VE19).} (T=[1-2]), decreasing to around 24\% (T=[2-3]) and increasing again to achieve 40\% at the end of the simulation (T=[5-6]). The $A\lambda04$ model has a larger number of B-type orbits at the bar saturation phase showing a maximum of 32\% of the disk mass fraction. Similar to the $A\lambda03$ model, the number of B-type orbits decreases after the bar saturation phase and later increases approximately up to 35\% of the disk mass fraction at the end of the simulation. The $A\lambda05$ model gets less than 30\% of B-type orbits at the bar saturation phase (T=[4-5]) and later decreases drastically to around 12\% of the disk mass fraction (T=[5-6]). To gain further insights, we continue monitoring the simulation of this model for an additional Gyr, aiming to determine whether the bar structure is ultimately destroyed or if it regrows \citep{2002A&A...392...83B}. In Figure \ref{fig:Nwrwy}, we observe that the fill circles, which represent model $A\lambda05$, extends up to 7 Gyr. We note that the disk mass fraction of B-type orbits increase in this last Gyr up to around 16\%.

Furthermore, the middle panel of Figure \ref{fig:Nwrwy} illustrates the evolution of the ratio $2.4<\omega_R/\omega_x<2.6$, revealing a significant increase as the saturation phase of the bar concludes. We observe an approximately 10\% increase for the $A\lambda03$ model, a 17\% increase for the $A\lambda04$ model, and a similar growth pattern in this type of orbit for the $A\lambda05$ model. On the other hand, in the lower panel, corresponding to orbits with a frequency ratio of $2.9<\omega_R/\omega_x<3.1$, we observe a slight increase during the saturation phase of the bar for the $A\lambda03$ and $A\lambda04$ models. However, this increase is more pronounced in the case of the $A\lambda05$ model, indicating a slower formation of the bar in this configuration.

Now, we focus on the frequency ratio $\omega_z/\omega_x$ to identify the orbital evolution which leads to the formation of the peanut structure. Figures \ref{fig:frozenL03}, \ref{fig:frozenL04}, and \ref{fig:frozenL05} show the number of B-type disk particles ($1.9<\omega_R/\omega_x<2.1$) as a function of $\omega_z/\omega_x$ ratio measured in different intervals of time. The left column shows the frequencies measured from the N-body model. The middle and right columns show the frequency ratio $\omega_z/\omega_x$ for the frozen potentials over 1 and 6 Gyr of integration time. In general, we notice that the shape of the curves of the frozen potentials results are very similar to the results of the N-body models. For instance, the highest peaks contribution appears in the same ratio frequency as the measures calculated in the N-body model, which means that the ratio frequency obtained from the N-body simulation for an integration time of 1 Gyr is enough to get a good approximation of the frequencies of the orbits.

The $A\lambda03$ model, depicted in Figure \ref{fig:frozenL03}, shows that the bar structure is formed by particles spread in the frequency ratio $2.0<\omega_z/\omega_x<3.5$ during the earliest times (top left panel). Thus, it corresponds to the initial stages of bar formation before the buckling phase occurs (see Figure \ref{fig:models}). Similarly, orbit scattering occurs for calculations of frozen potential by 1 Gyr (top middle panel); however, orbits with higher frequency ratios are lost when calculating the frozen potential by 6 Gyr. The aforementioned effect could happen because it is challenging to maintain the frequency of orbits for more than 1 Gyr in the potential where the bar is still weak. Later, when the bar is fully established, and the orbits start to swing from the disk plane (the buckling phase begins), all the stellar orbits fall within the frequency ratio interval $1.4<\omega_z/\omega_x<2.2$ from the second to the sixth Gyr.

We can relate Figure \ref{fig:frozenL03} with the classification developed by \cite{2015MNRAS.450L..66P}. They classified the orbits according to the frequencies ratios and the shapes using the letters A, B, C, D, E, and F for the intervals $1.45<\omega_z/\omega_x<1.55$,  $1.55<\omega_z/\omega_x<1.65$, $1.65<\omega_z/\omega_x<1.75$, $1.75<\omega_z/\omega_x<1.85$, $1.85<\omega_z/\omega_x<1.95$, and $1.95<\omega_z/\omega_x<2.05$, respectively. Therefore, it is possible to identify these frequency intervals in the histograms of Figure 7 for the N-body model (left column) and its frozen potential (middle and right columns). In this manner, the first peak coincides with a frequency ratio from 1.4 to 1.5, which aligns with the A classification of \cite{2015MNRAS.450L..66P}. This growth is throughout the evolution of the bar structure. A second peak, which is close the frequency ratio $\omega_z/\omega_x=1.8$ from the fourth to sixth Gyr, corresponds to the D classification. And the last peak corresponds to the F classification ($\omega_z/\omega_x=2.0$); this peak keeps almost equal during the evolution of the bar structure. We can observe that during the buckling phase, around the third Gyr, a considerable number of F and A orbits appears. Then, the peak for the F orbits is more or less constant, and the first and second peaks increase (A and D orbits) as the bar evolves. The $1.4<\omega_z/\omega_x<1.5$ interval (D orbits) has the highest peak. Furthermore, all other types of orbits increase {the} contribution to the bar structure in some extent. It can be {noted} unlike the N-body model, in the calculations made for the frozen potential integrated by 6 Gyr, the peak located in the ratio 1.4 has a greater contribution than the other frequency intervals, which corresponds to orbits populating in the center of the galaxy, (see Figure \ref{fig:L03_map}).

Figure \ref{fig:frozenL04} shows the frequency ratio $\omega_z/\omega_x$  evolution for the $A\lambda04$ model. As we can observe, in this case, the outcomes of frequency measures from the frozen potential are very similar to the N-body model, where the form and the contribution of the ratio of frequencies appear in the same intervals (see middle and right columns of this figure). At earlier stages of the bar formation, we observe that our B-type orbits have a histogram similar to the one from model $A\lambda03$, but delayed in time and have larger $\omega_z/\omega_x$ (compare T=[1-2] for model $A\lambda03$ and T=[2-3] for model $A\lambda04$). In the next Gyr interval, the histogram for model $A\lambda04$ moves to lower $\omega_z/\omega_x$. Then, when the bar begins to develop the buckling phase, the orbits are established in the frequency ratio interval $1.4<\omega_z/\omega_x<2.2$, similar to the $A\lambda03$ model. At the end of the simulation, a high peak in the $1.7<\omega_z/\omega_x<1.9$ ratio appears, which aligns with the D classification, according to \cite{2015MNRAS.450L..66P}.

The halo-dominated model ($A\lambda05$) shows quite different behavior. {In} the first instance, a much lower number of disk particles stand in the $1.9<\omega_R/\omega_x<2.1$ range (B-type orbits). That is why the maximum scale in the left column of Figure \ref{fig:frozenL05} reaches only half of the value plotted for the other two models (see Figures \ref{fig:frozenL03} and \ref{fig:frozenL04}). At the early stages of the bar formation, the orbits locate around the frequency ratio $\omega_z/\omega_x\approx3$. These high $\omega_z/\omega_x$ values represent a morphology of a boxy structure. Similar to the $A\lambda03$ and $A\lambda04$ model, the histogram moves toward the left as the bar evolves, but we notice that the $A\lambda05$ model does not enter the buckling phase. For this model, it is only at the end of the simulation when particles participating in the bar structure present a distribution with lower $\omega_z/\omega_x$; the orbits are beginning to oscillate outside the disk plane with lower $\omega_z/\omega_x$, and only then they will be able to evolve from a boxy to a peanut shape. We should note that the results from the frozen potential of this model are also quite similar to the N-body model: the histograms appear in the same quotient of frequencies, and they have very similar behavior.

We have performed a similar analysis in the 
\lstinline[language=Fortran]{GravPot16}
potential that resembles the main complexities of the Milky-Way galaxy (Chaves-Velasquez et al., in preparation). This potential has been widely used in works related to the dynamics of the Galaxy \citep[see for instance,][]{2020MNRAS.495.4113F, 2021A&A...647A..64F, 2021ApJ...918L..37F}. Our histograms show distributions of frequencies at high values for the ratio $\omega_{z}/\omega_{x}$ coefficient, similar to models $A\lambda 03$ and $A\lambda 04$ at the early stages of the simulation before the galactic buckling. We have studied the orbital families that shape the B/P morphology, and we found that orbits at higher vertical bifurcations of the planar x1 family \citep[x1v3, x1v5, x1v7 according to the notation of][]{2002MNRAS.333..861S} contribute to shaping the edge-on view of the bar. 

\subsection{Evolution of the orbital assembly}\label{sec:eoa}

\begin{figure*}
	\includegraphics[scale=0.8]{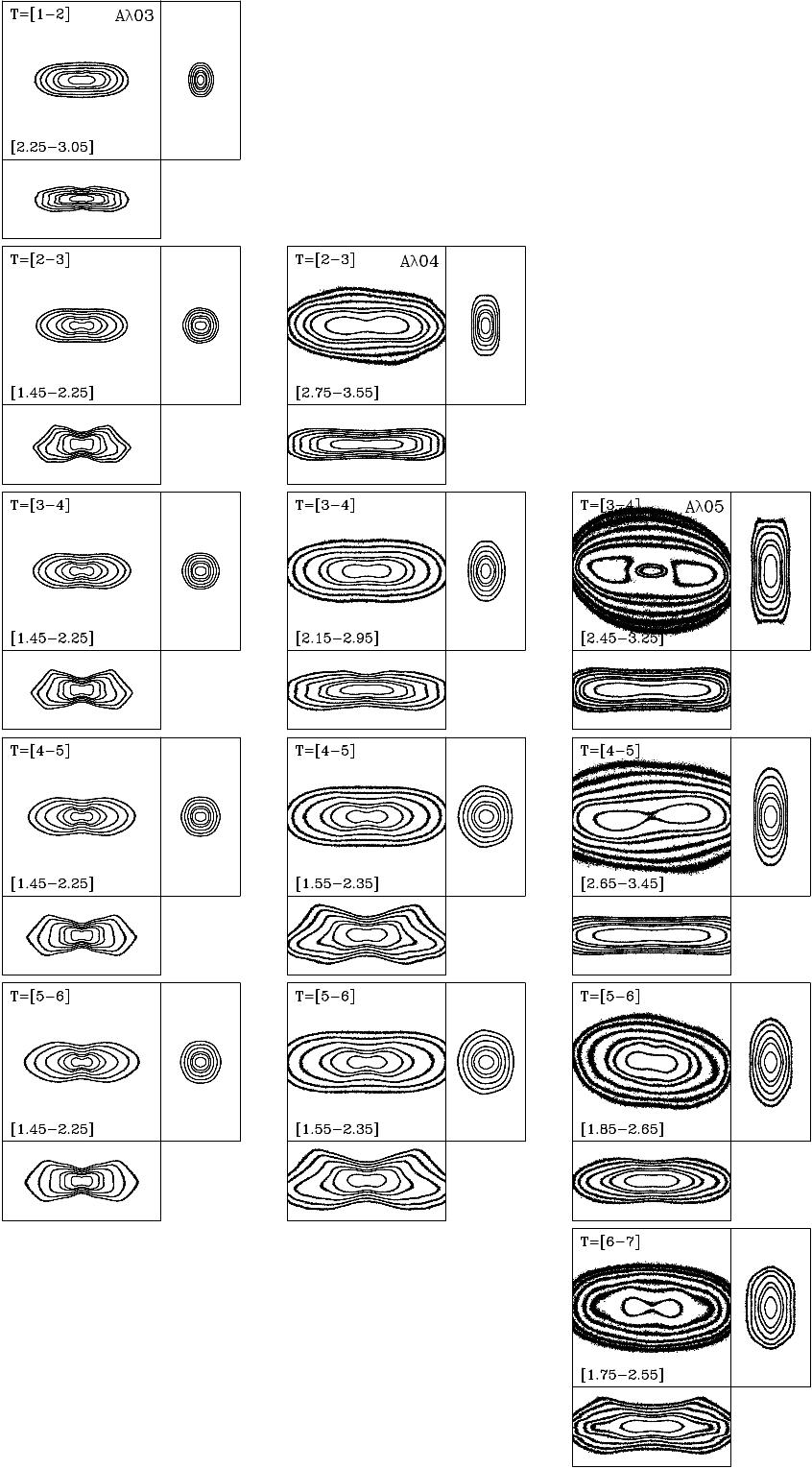}
	\caption{Orbits assembly density profile (OADP) maps in face-on, edge-on, and end-on views for particles with ratio $1.9<\omega_R/\omega_x<2.1$ and $\omega_z/\omega_x$ as indicated in the bottom left of the face-on views. The left, middle, and right panels show the models $A\lambda03$, $A\lambda04$, and $A\lambda05$, respectively. The time increases downwards. Each panel represents a different model at different time intervals and a different range of the $\omega_z/\omega_x$ ratio (see left column of Figures \ref{fig:frozenL03}, \ref{fig:frozenL04}, and \ref{fig:frozenL05}). The x and y axes are from $-5.0$ to $5.0$ kpc, and the vertical direction z, from $-2.5$ to $2.5$ kpc. Each map is normalized to its maximum, and six contours are plotted (0.025, 0.05, 0.1, 0.2, 0.3 and 0.6). We can easily follow the buckling formation in the $A\lambda03$ model. In the beginning, it is quite asymmetric concerning the disk plane, and later on, this structure becomes symmetric. Instead, at the end of our simulation, $A\lambda04$ does not fully achieve the buckling process. In the case of $A\lambda05$, the full bar formation process is delayed, and the bisymmetric structure is just incipient at the end of the simulation.}
	\label{fig:maps}
\end{figure*}

We now discuss the morphologies that arise from assembling 2:1 resonance orbits considering their face-on, edge-on, and end-on views and their characteristic frequency ratios. To distinguish the contribution to the shape of the bar structure, Figure \ref{fig:maps} shows orbits assembly density profile (OADP) in their different views for B-type orbits of different $\omega_z/\omega_x$; the range of $\omega_z/\omega_x$, which is used to construct these maps, are displayed in the histograms of Figures \ref{fig:frozenL03}, \ref{fig:frozenL04}, and \ref{fig:frozenL05} (thick vertical lines). In Figure \ref{fig:maps}, the left, middle, and right columns represent the $A\lambda03$, $A\lambda04$, and $A\lambda05$ models, respectively. The range of $\omega_z/\omega_x$ used for each model at each time interval is written in the bottom left of the XY views.

We observe that at the early stages of the bar formation, the orbits that participate in the oval/boxy distortion have high values of $\omega_z/\omega_x$. The disk-dominated model develops mainly an oval shape, and the other models display a more boxy morphology. The models arrive at this stage at different times; T=[1-2] Gyr for the disk-dominated model,  T=[2-3] Gyr for the intermediate one, and T=[3-4] Gyr for the halo-dominated model. 

The next stage in the evolution of the bar structure is the buckling phase. This phase can also be recognized in Figures \ref{fig:frozenL03}, \ref{fig:frozenL04}, and \ref{fig:frozenL05}. The histograms for the frequency ratio $\omega_z/\omega_x$ move towards lower values of this ratio. That means the particles will oscillate less in the vertical direction while undergoing the oscillation following the bar potential. The $A\lambda03$ and $A\lambda04$ models evolve to that phase at T=[2-3] Gyr and T=[4-5] Gyr, respectively. For the halo-dominated $A\lambda05$ model, this phase is not yet fully achieved at the end of the time simulation (T=7 Gyr). 

In Appendix \ref{appendixa}, we present the OADP maps for our models in bins of 0.1*$\omega_z/\omega_x$ for the reader better understand the contribution of different kinds of orbits to the morphology of the bar structure.

\subsection{Orbit shapes}

\begin{figure*}
	\centering
	\includegraphics[scale=0.85,angle=-90]{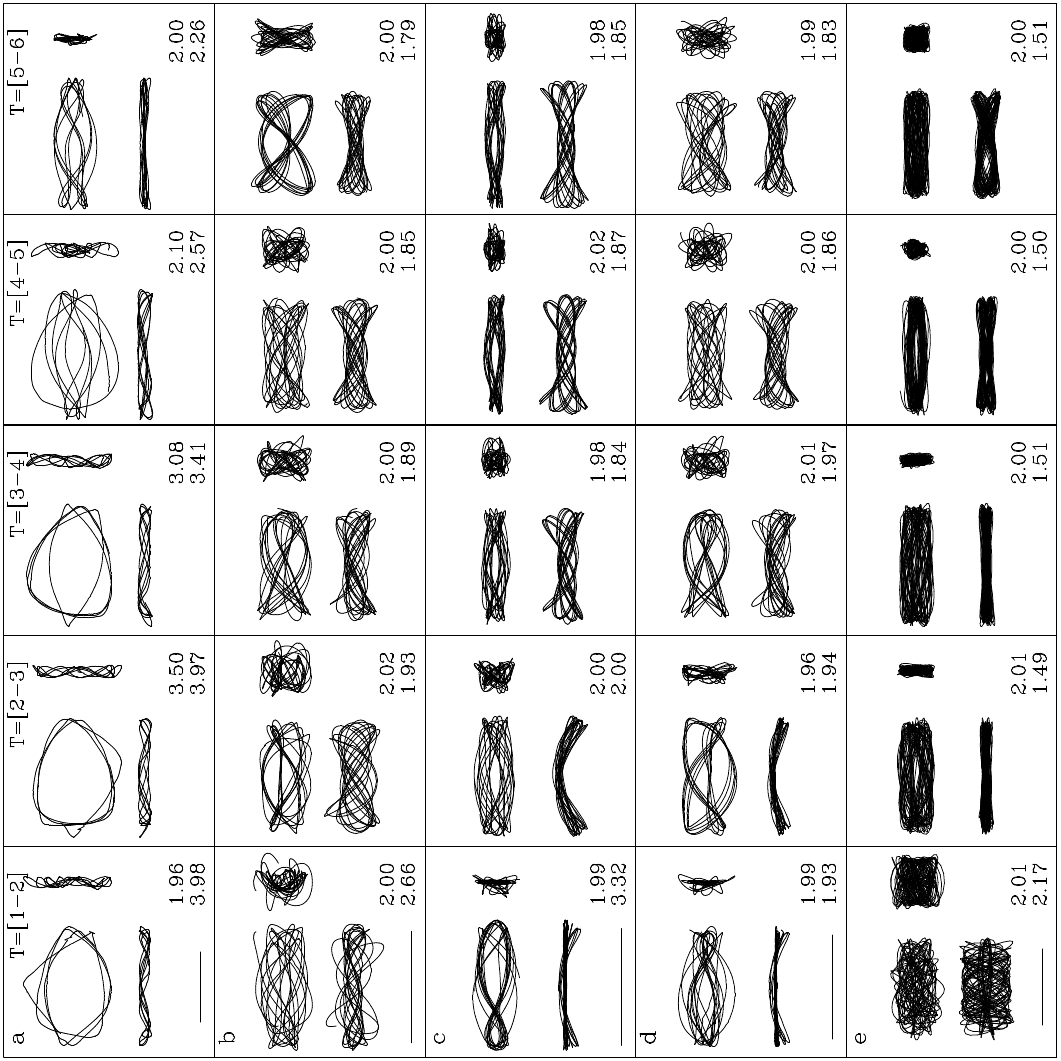}
	\caption{Evolution of five orbits in their different views from the $A\lambda03$ model is presented. For all the particles, the time increases to the right in units of one Gyr. The horizontal line in the bottom left location in the left panels represents 5.0 kpc for particles a) to d). For particle e), the horizontal line represents 1.0 kpc. The numbers at the bottom right of each panel are the values for $\omega_R/\omega_x$ (upper value) and the $\omega_z/\omega_x$ (bottom value).}
	\label{fig:L03_orbs_B_shape}
\end{figure*}

We presented in Figures \ref{fig:frozenL03}, \ref{fig:frozenL04}, and \ref{fig:frozenL05} the evolution of the $\omega_z/\omega_x$ ratio distribution, which means that the orbits populated different values of this ratio during the growth of the bar structure. To illustrate this effect, in Figure \ref{fig:L03_orbs_B_shape}, we show the evolution of five orbits for the N-body model $A\lambda03$ which are B-type ($1.9<\omega_R/\omega_x<2.1$) at least in the last time interval of analysis (T=[5-6] Gyr). In that figure, we show xy, xz, and yz views of each orbit for five different time intervals. We can recognize several orbits morphology, which appears in models using a Ferrers analytical bar \citep[e.g.,][among others]{2002MNRAS.333..847S,2002MNRAS.333..861S,2002MNRAS.337..578P}.

The `a' orbit of Figure \ref{fig:L03_orbs_B_shape} is swinging on the bar region between $2.3<\omega_z/\omega_x<4.0$ to finally be part of the bar 2:1 resonance at the last time interval. We can perceive that this type of orbit contributes to the bar structure, but not to the peanut shape. It arrives later to the 2:1 resonance and stays on the disk plane. Likewise, we notice that these types of orbits do not add so much to the bar potential because the contribution of these orbits is low (see Figure \ref{fig:frozenL03}). 

The `b' orbit is part of the B-type in the whole of its evolution. This orbit first appears on $\omega_z/\omega_x=2.66$ frequency ratio.
Next, it moves to $\omega_z/\omega_x=1.93$ at T=[2-3].
Finally, for the following time intervals, this orbit contributes to the peanut shape of the bar in the edge-on view. Moreover, at the last time interval, this orbit shows a pretzel shape in the face-on view, a peanut shape in the edge-on view, and an X shape in the end-on view. Besides, this orbit contributes to the bump of the distribution located in $\omega_z/\omega_x\approx2.0$ in Figure \ref{fig:frozenL03}. This orbit contributes to the bin located at about $\omega_z/\omega_x\approx1.8$ in the rest of the simulation.

The $\omega_z/\omega_x$ ratio in the `c' orbit decreases from 3.32 to 2.00, contributing to the buckling of the bar (first two panels, T=[1-2] and T=[2-3]). This orbit shows a banana shape in the edge-on view. Then, the $\omega_z/\omega_x$ ratio decreases to 1.85 and remains almost constant during the evolution.
At the last time interval, this orbit shows a peanut shape in the edge-on view and an elongated rectangular shape in the face-on and end-on views.

For the `d' trajectory, the $\omega_z/\omega_x$ frequency ratio experiences a slight reduction from 1.93 to 1.83. Initially, it closely aligns with the bin distribution around $\omega_z/\omega_x\approx2.0$. This orbit exhibits a clear banana shape in the xz view and a pretzel shape in the xy view (T=[2-3]). Over time, the pretzel shape evolves into a fish shape in the face-on view. Eventually, the orbit displays a boxy contour in the face-on view, a peanut shape in the edge-on view, and an oval shape in the end-on view.

The fifth orbit contributes to the central part of the bar; the  $\omega_z/\omega_x$ ratio moves from 2.17  to 1.51 during the evolution. This orbit has an elongated shape in the face-on and edge-on view. At the last time-interval, the orbit depicts a boxy/peanut shape in the end-on view. These kinds of orbits contribute to the central boxy structure of the model (see Figure \ref{fig:maps}).

Considering the previous description of the orbits in Figure \ref{fig:L03_orbs_B_shape}, it can be said that the particles change their frequencies during the evolution of the model. In general, the $\omega_z/\omega_x$ ratio decreases from high values before the buckling phase to lower values after the buckling (see Figure \ref{fig:frozenL03}). This effect happens earlier for disk-dominated models. The orbits change their morphologies as they evolve with the bar potential; therefore, different orbital families in the models, which form the backbone of the bar structure in a disk galaxy, are assembled by distinct particles during the evolution. We must note that this effect is not exhibited when we study orbits using Ferrers analytical potentials or when we `freeze' a simulation, calculate the potential produced by the particle distribution, and calculate orbits in this numerical potential. These techniques do not permit the self-consistent evolution of the bar potential. 

\section{Discussion}

The present work extends the studies by \cite{2015MNRAS.450L..66P} and \cite{2016ApJ...830..108G}. In \cite{2015MNRAS.450L..66P}, they performed frequency analysis on orbits calculated in frozen potentials. On the other hand, \cite{2016ApJ...830..108G} used orbits taken from a fully self-consistent N-body simulation, and the bar, in their model, was tidally induced. They present the frequency analysis in a single interval of around 1.6 Gyrs. During this interval, the bar in their model exhibits a constant pattern speed.

In this work, we studied the orbital structure evolution for three N-body models from VE19 using the dominant frequencies of the disk particles. The models differ on the values of the initial Spin parameter of the disk $\lambda_d$. We used simulations with high time resolution, and then we could analyze the orbital frequency distribution of the bar structure in a disk galaxy as the bar potential evolves. In the process of bar formation, the main resonances 2:1, 5:2, and 3:1 are populated \citep{2007MNRAS.381..757V}. Our disk-dominated model replenishes these resonances at earlier stages of the model evolution, while it gets much longer for a halo-dominated model. In the present study, we focused on B-type orbits, which fill the 2:1 resonance ($1.9<\omega_R/\omega_x<2.1$).

On the frequency ratio plots $\Omega/\omega_z$ vs. $\omega_R/\omega_z$ (Figure \ref{fig:all_MAP_ok_B}), we can perceive how the main resonances replenish as the models evolve. Once the bar forms, a large number of particles occupy the 2:1 resonance, with a lower contribution of orbits with frequencies between the 5:2 and 3:1 resonances (see Figure \ref{fig:Nwrwy}). For this work, we focused our analysis on the 2:1 resonance (B-type orbits).

We show that the distribution of B-type orbits in the $\omega_z/\omega_x$ ratio differs among all models at different time intervals. During the bar formation in the disk-dominated model (T=[1-2]), the B-type orbits exhibit a ratio interval of $2<\omega_z/\omega_x<3$. In the subsequent Gyrs, as the bar reaches its mature state (boxy/peanut shape and buckling), the B-type orbits show a ratio interval of $1.5<\omega_z/\omega_x<2.2$ (see Figures \ref{fig:frozenL03}, \ref{fig:frozenL04}, and \ref{fig:frozenL05}). In the intermediate $A\lambda04$ model, high values for $\omega_z/\omega_x$ are achieved only in the interval T=[2-3]. The orbits transition to lower $\omega_z/\omega_x$ much slower than the disk-dominated model and reach the ratio interval of $1.5<\omega_z/\omega_x<2.2$ only after the fourth Gyr. At that moment, the orbits begin to leave the disk plane; in other words, the buckling phase initiates. Finally, the halo-dominant $A\lambda05$ model evolves in a significantly different manner. The B-type orbits exhibit high values for the ratio $\omega_z/\omega_x$ throughout the entire simulation (Figure \ref{fig:frozenL05}). In this case, the distribution gradually shifts to the left as the model evolves from the fifth to the seventh Gyr. In this halo-dominated model, the buckling of the bar is just incipient at the end of our simulation (T=7 Gyr).

As we can observe, the frequency ratios drift as the bar evolves. These shifts are more noticeable during the buckling phase of the bar. After this event, some orbits remain within a specific range of these frequency ratios (see the bins in Figures \ref{fig:frozenL03}, \ref{fig:frozenL04} or \ref{fig:frozenL05}, and the orbits in Figure \ref{fig:L03_orbs_B_shape}). These particular orbits can be potentially categorized as regular or quasiperiodic orbits due to their minimal deviations, as demonstrated by \cite{2016MNRAS.463.3499W}. In that study, the orbits were classified into regular and chaotic orbits using the frequency drift technique, revealing that this classification depends on the orbit projection, where the orbit appears to be more regular in the xy view than in the other perspectives. Hence, it can infer that orbits persisting within the same bin of the distributions over time can be classified as regular orbits.

Furthermore, consistent findings are obtained when employing the frozen potential. The frequency distribution exhibits a similar pattern to that observed in the N-body simulations. However, for a more comprehensive elucidation of the dynamics of models, an alternative approach is available whereby orbits can be computed using a frozen potential derived from Basis Function Expansion \citep[e.g.][]{1992ApJ...386..375H,1995MNRAS.277.1341K,2015MNRAS.450.2842V,2018MNRAS.476.2092L,2019MNRAS.482.1525V,2020A&A...639A..38W,2022A&A...668A..55W}.

In a recent study by \cite{2022A&A...668A..55W}, they demonstrated that there exist periodic orbits exhibiting diverse pairs of morphological families with multiplicities{\footnote{Multiplicity represents the number of revolutions needed for an orbit to close in phase space.}} of at least two ($\mathscr{M}=2$) or more. Notably, certain orbits, such as the pretzel and fish orbits, exhibit a multiplicity of two. Therefore, in N-body simulations, the orbits supporting the bar structure can manifest distinct multiplicities of quasi-periodic types. As a consequence, it changes as the bar potential evolves, like we illustrate in Figure 11. For instance, orbit `d' demonstrates two different $\mathscr{M}=2$ shapes during two investigated time intervals. Furthermore, it is plausible to consider a correlation between the multiplicity, and the frequencies of the orbits. Thus, an extension of this study is warranted to explore the relationship between multiplicity and orbit frequencies in greater detail.

We also plot the orbits assembly density profile (OADP) maps to study the morphology of the B-type orbits at different $\omega_z/\omega_x$ intervals (Figures \ref{fig:maps}, \ref{fig:L03_map}, \ref{fig:L04_map} and \ref{fig:L05_map}). 
In the $A\lambda03$ and $A\lambda04$ models, the orbits located at the interval $1.5<\omega_z/\omega_x<2.2$ are pulled out from the disk plane. The orbits showing lower values of $\omega_z/\omega_x$ are located in the inner part of the disk and are responsible for the internal boxy/X shape. The orbits having larger $\omega_z/\omega_x$ participated in the evolution of the buckling structure. These orbits buckle first showing a `smile down' figure \citep{2016ApJ...830..108G} and then evolve to a peanut shape (see Figure \ref{fig:L03_orbs_B_shape}, orbit `d'). The halo-dominated model $A\lambda05$ develops a bar structure much later compared to the other models, and the buckling phase is just starting at the end of the simulation.

In \cite{erwin2013peanuts}, the authors present several moderately inclined galaxies as visual examples of boxy bars. The study highlights that galaxies such as NGC 5377 and IC5240 exhibit strong bars with broad and distinctly boxy shapes in their isophotes, resembling the $A\lambda03$ and $A\lambda04$ models. Additionally, they identified other galaxies like NGC 3049 and IC 676, which possess bars that have not vertically thickened, similar to the $A\lambda05$ model. Roughly speaking, based on our findings, we can conclude that the boxy/peanut-shaped bar structures for these observed galaxies are supported by a frequency range of $1.45<\omega_z/\omega_x<2.55$, with smaller ratios predominantly located in the central region of the bar, while larger quotients find in the peripheral region. Conversely, bars in galaxies exhibiting any boxy shape sustain likely frequency ranges $\omega_z/\omega_x>2.55$.

The higher frequencies maintained in the disk plane are probably related to the velocity dispersions. A lower velocity dispersion indicates stronger interactions among particles compared to those with a higher velocity dispersion. The stronger interactions experienced by particles with low velocity dispersion cause them to deviate from the disk plane. This instability in the disk-spin parameter results in particles orbiting out of the disk plane with lower $\omega_z$ frequency (VE19). The calculation of the evolution of a set of models with different initial vertical velocity dispersions should bring light to this subject.

\section{Summary}

In summary, the frequency analysis of the N-body models reveals a distinction in the vertical behavior of the orbits. The models, which are dominated by the disk component, catch around 30\% of disk particles in the 2:1 resonance. During the initial stages of bar formation, prior to the buckling phase, large ratios of $\omega_z/\omega_x$ are observed. Later on in the evolution of the models, the distribution of $\omega_z/\omega_x$ moves to lower values. The buckling process is fully evolved in the $A\lambda03$ model, partially developed in the $A\lambda04$ one, and just beginning in the $A\lambda05$ case within our models. The main results of this study are as follows:


\begin{itemize}
    \item The orbits that form the bar structure prior to the buckling phase are within the range of $2.25<\omega_z/\omega_x<3.05$ for the $A\lambda03$ model, which has a smaller scale radius, and $2.75<\omega_z/\omega_x<3.55$ for the $A\lambda04$ and $A\lambda05$ models, which have a larger scale radius.

    \item The orbits that form the bar structure after the buckling phase transition are within the range of $1.45<\omega_z/\omega_x<2.25$ for the $A\lambda03$ model and $1.55<\omega_z/\omega_x<2.35$ for the $A\lambda04$ model. The $A\lambda05$ model shows a weak buckling in the range of $1.75<\omega_z/\omega_x<2.55$ during the final time interval.

    \item Lastly, we observe that after the buckling phase, the orbits with smaller ratios of $\omega_z/\omega_x$ are located in the central region of the bar structure, and these orbits contribute the most to the potential of the bar. On the other hand, orbits with larger ratios of $\omega_z/\omega_x$ are found in the peripheral regions of the bar (see Figures in Appendix \ref{appendixa}).
\end{itemize}

\section*{DATA AVAILABILITY}

The data underlying this paper will be shared on reasonable request to the corresponding author.

\section*{Acknowledgements}


We thank the referees and the scientific editor for comments and suggestions that greatly improved the manuscript.
D.V.E thanks the Facultad de Ingenier\'\i a and Direcci\'on de Investigaciones of Universidad Mariana, Colombia  (Project 239). L.C.V. acknowledges the support of the postdoctoral Fellowship of DGAPA-UNAM, Mexico. I.P. thanks the Mexican Foundation Conahcyt.



\bibliographystyle{mnras}
\bibliography{example} 

\begin{thebibliography}{}
\makeatletter
\relax
\def\mn@urlcharsother{\let\do\@makeother \do\$\do\&\do\#\do\^\do\_\do\%\do\~}
\def\mn@doi{\begingroup\mn@urlcharsother \@ifnextchar [ {\mn@doi@}
  {\mn@doi@[]}}
\def\mn@doi@[#1]#2{\def\@tempa{#1}\ifx\@tempa\@empty \href
  {http://dx.doi.org/#2} {doi:#2}\else \href {http://dx.doi.org/#2} {#1}\fi
  \endgroup}
\def\mn@eprint#1#2{\mn@eprint@#1:#2::\@nil}
\def\mn@eprint@arXiv#1{\href {http://arxiv.org/abs/#1} {{\tt arXiv:#1}}}
\def\mn@eprint@dblp#1{\href {http://dblp.uni-trier.de/rec/bibtex/#1.xml}
  {dblp:#1}}
\def\mn@eprint@#1:#2:#3:#4\@nil{\def\@tempa {#1}\def\@tempb {#2}\def\@tempc
  {#3}\ifx \@tempc \@empty \let \@tempc \@tempb \let \@tempb \@tempa \fi \ifx
  \@tempb \@empty \def\@tempb {arXiv}\fi \@ifundefined
  {mn@eprint@\@tempb}{\@tempb:\@tempc}{\expandafter \expandafter \csname
  mn@eprint@\@tempb\endcsname \expandafter{\@tempc}}}

\bibitem[\protect\citeauthoryear{{Abbott}, {Valluri}, {Shen}  \&
  {Debattista}}{{Abbott} et~al.}{2017}]{2017MNRAS.470.1526A}
{Abbott} C.~G.,  {Valluri} M.,  {Shen} J.,   {Debattista} V.~P.,  2017, \mn@doi
  [\mnras] {10.1093/mnras/stx1262}, \href
  {https://ui.adsabs.harvard.edu/abs/2017MNRAS.470.1526A} {470, 1526}

\bibitem[\protect\citeauthoryear{{Aguerri}, {M{\'e}ndez-Abreu}  \&
  {Corsini}}{{Aguerri} et~al.}{2009}]{2009A&A...495..491A}
{Aguerri} J.~A.~L.,  {M{\'e}ndez-Abreu} J.,   {Corsini} E.~M.,  2009, \mn@doi
  [\aap] {10.1051/0004-6361:200810931}, \href
  {https://ui.adsabs.harvard.edu/abs/2009A&A...495..491A} {495, 491}

\bibitem[\protect\citeauthoryear{{Athanassoula}}{{Athanassoula}}{2002}]{2002ApJ...569L..83A}
{Athanassoula} E.,  2002, \mn@doi [\apjl] {10.1086/340784}, \href
  {https://ui.adsabs.harvard.edu/abs/2002ApJ...569L..83A} {569, L83}

\bibitem[\protect\citeauthoryear{{Athanassoula}}{{Athanassoula}}{2003}]{2003MNRAS.341.1179A}
{Athanassoula} E.,  2003, \mn@doi [\mnras] {10.1046/j.1365-8711.2003.06473.x},
  \href {https://ui.adsabs.harvard.edu/abs/2003MNRAS.341.1179A} {341, 1179}

\bibitem[\protect\citeauthoryear{Athanassoula}{Athanassoula}{2013}]{2013seg..book..305A}
Athanassoula E.,  2013, in Falc{\'o}n-Barroso J.,  Knapen J.~H.,  eds, ,
  Secular Evolution of Galaxies.
Cambridge University Press, Cambridge, UK, p.~305

\bibitem[\protect\citeauthoryear{{Athanassoula}, {Bienayme}, {Martinet}  \&
  {Pfenniger}}{{Athanassoula} et~al.}{1983}]{1983A&A...127..349A}
{Athanassoula} E.,  {Bienayme} O.,  {Martinet} L.,   {Pfenniger} D.,  1983,
  \aap, \href {https://ui.adsabs.harvard.edu/abs/1983A&A...127..349A} {127,
  349}

\bibitem[\protect\citeauthoryear{{Athanassoula}, {Machado}  \&
  {Rodionov}}{{Athanassoula} et~al.}{2013}]{2013MNRAS.429.1949A}
{Athanassoula} E.,  {Machado} R. E.~G.,   {Rodionov} S.~A.,  2013, \mn@doi
  [\mnras] {10.1093/mnras/sts452}, \href
  {https://ui.adsabs.harvard.edu/abs/2013MNRAS.429.1949A} {429, 1949}

\bibitem[\protect\citeauthoryear{{Bournaud} \& {Combes}}{{Bournaud} \&
  {Combes}}{2002}]{2002A&A...392...83B}
{Bournaud} F.,  {Combes} F.,  2002, \mn@doi [\aap]
  {10.1051/0004-6361:20020920}, \href
  {https://ui.adsabs.harvard.edu/abs/2002A&A...392...83B} {392, 83}

\bibitem[\protect\citeauthoryear{{Buta} et~al.,}{{Buta}
  et~al.}{2015}]{2015ApJS..217...32B}
{Buta} R.~J.,  et~al., 2015, \mn@doi [\apjs] {10.1088/0067-0049/217/2/32},
  \href {https://ui.adsabs.harvard.edu/abs/2015ApJS..217...32B} {217, 32}

\bibitem[\protect\citeauthoryear{{Carpintero} \& {Aguilar}}{{Carpintero} \&
  {Aguilar}}{1998}]{1998MNRAS.298....1C}
{Carpintero} D.~D.,  {Aguilar} L.~A.,  1998, \mn@doi [\mnras]
  {10.1046/j.1365-8711.1998.01320.x}, \href
  {https://ui.adsabs.harvard.edu/abs/1998MNRAS.298....1C} {298, 1}

\bibitem[\protect\citeauthoryear{{Ceverino} \& {Klypin}}{{Ceverino} \&
  {Klypin}}{2007}]{2007MNRAS.379.1155C}
{Ceverino} D.,  {Klypin} A.,  2007, \mn@doi [\mnras]
  {10.1111/j.1365-2966.2007.12001.x}, \href
  {https://ui.adsabs.harvard.edu/abs/2007MNRAS.379.1155C} {379, 1155}

\bibitem[\protect\citeauthoryear{{Chaves-Velasquez}, {Patsis}, {Puerari},
  {Skokos}  \& {Manos}}{{Chaves-Velasquez} et~al.}{2017}]{2017ApJ...850..145C}
{Chaves-Velasquez} L.,  {Patsis} P.~A.,  {Puerari} I.,  {Skokos} C.,   {Manos}
  T.,  2017, \mn@doi [\apj] {10.3847/1538-4357/aa961a}, \href
  {https://ui.adsabs.harvard.edu/abs/2017ApJ...850..145C} {850, 145}

\bibitem[\protect\citeauthoryear{{Combes} \& {Sanders}}{{Combes} \&
  {Sanders}}{1981}]{1981A&A....96..164C}
{Combes} F.,  {Sanders} R.~H.,  1981, \aap, \href
  {https://ui.adsabs.harvard.edu/abs/1981A&A....96..164C} {96, 164}

\bibitem[\protect\citeauthoryear{{Contopoulos} \& {Harsoula}}{{Contopoulos} \&
  {Harsoula}}{2013}]{2013MNRAS.436.1201C}
{Contopoulos} G.,  {Harsoula} M.,  2013, \mn@doi [\mnras]
  {10.1093/mnras/stt1640}, \href
  {https://ui.adsabs.harvard.edu/abs/2013MNRAS.436.1201C} {436, 1201}

\bibitem[\protect\citeauthoryear{{Contopoulos} \&
  {Papayannopoulos}}{{Contopoulos} \&
  {Papayannopoulos}}{1980}]{1980A&A....92...33C}
{Contopoulos} G.,  {Papayannopoulos} T.,  1980, \aap, \href
  {https://ui.adsabs.harvard.edu/abs/1980A&A....92...33C} {92, 33}

\bibitem[\protect\citeauthoryear{{Debattista}, {Carollo}, {Mayer}  \&
  {Moore}}{{Debattista} et~al.}{2004}]{2004ApJ...604L..93D}
{Debattista} V.~P.,  {Carollo} C.~M.,  {Mayer} L.,   {Moore} B.,  2004, \mn@doi
  [\apjl] {10.1086/386332}, \href
  {https://ui.adsabs.harvard.edu/abs/2004ApJ...604L..93D} {604, L93}

\bibitem[\protect\citeauthoryear{{Debattista}, {Mayer}, {Carollo}, {Moore},
  {Wadsley}  \& {Quinn}}{{Debattista} et~al.}{2006}]{2006ApJ...645..209D}
{Debattista} V.~P.,  {Mayer} L.,  {Carollo} C.~M.,  {Moore} B.,  {Wadsley} J.,
   {Quinn} T.,  2006, \mn@doi [\apj] {10.1086/504147}, \href
  {https://ui.adsabs.harvard.edu/abs/2006ApJ...645..209D} {645, 209}

\bibitem[\protect\citeauthoryear{Erwin \& Debattista}{Erwin \&
  Debattista}{2013}]{erwin2013peanuts}
Erwin P.,  Debattista V.~P.,  2013, Monthly Notices of the Royal Astronomical
  Society, 431, 3060

\bibitem[\protect\citeauthoryear{{Erwin} \& {Debattista}}{{Erwin} \&
  {Debattista}}{2017}]{2017MNRAS.468.2058E}
{Erwin} P.,  {Debattista} V.~P.,  2017, \mn@doi [\mnras]
  {10.1093/mnras/stx620}, \href
  {https://ui.adsabs.harvard.edu/abs/2017MNRAS.468.2058E} {468, 2058}

\bibitem[\protect\citeauthoryear{{Eskridge} et~al.,}{{Eskridge}
  et~al.}{2000}]{2000AJ....119..536E}
{Eskridge} P.~B.,  et~al., 2000, \mn@doi [\aj] {10.1086/301203}, \href
  {http://adsabs.harvard.edu/abs/2000AJ....119..536E} {119, 536}

\bibitem[\protect\citeauthoryear{{Fern{\'a}ndez-Trincado}, {Chaves-Velasquez},
  {P{\'e}rez-Villegas}, {Vieira}, {Moreno}, {Ortigoza-Urdaneta}  \&
  {Vega-Neme}}{{Fern{\'a}ndez-Trincado} et~al.}{2020}]{2020MNRAS.495.4113F}
{Fern{\'a}ndez-Trincado} J.~G.,  {Chaves-Velasquez} L.,  {P{\'e}rez-Villegas}
  A.,  {Vieira} K.,  {Moreno} E.,  {Ortigoza-Urdaneta} M.,   {Vega-Neme} L.,
  2020, \mn@doi [\mnras] {10.1093/mnras/staa1386}, \href
  {https://ui.adsabs.harvard.edu/abs/2020MNRAS.495.4113F} {495, 4113}

\bibitem[\protect\citeauthoryear{{Fern{\'a}ndez-Trincado}
  et~al.,}{{Fern{\'a}ndez-Trincado} et~al.}{021a}]{2021A&A...647A..64F}
{Fern{\'a}ndez-Trincado} J.~G.,  et~al., 2021a, \mn@doi [\aap]
  {10.1051/0004-6361/202040255}, \href
  {https://ui.adsabs.harvard.edu/abs/2021A&A...647A..64F} {647, A64}

\bibitem[\protect\citeauthoryear{{Fern{\'a}ndez-Trincado}
  et~al.,}{{Fern{\'a}ndez-Trincado} et~al.}{021b}]{2021ApJ...918L..37F}
{Fern{\'a}ndez-Trincado} J.~G.,  et~al., 2021b, \mn@doi [\apjl]
  {10.3847/2041-8213/ac225b}, \href
  {https://ui.adsabs.harvard.edu/abs/2021ApJ...918L..37F} {918, L37}

\bibitem[\protect\citeauthoryear{{Gajda}, {{\L}okas}  \&
  {Athanassoula}}{{Gajda} et~al.}{2016}]{2016ApJ...830..108G}
{Gajda} G.,  {{\L}okas} E.~L.,   {Athanassoula} E.,  2016, \mn@doi [\apj]
  {10.3847/0004-637X/830/2/108}, \href
  {https://ui.adsabs.harvard.edu/abs/2016ApJ...830..108G} {830, 108}

\bibitem[\protect\citeauthoryear{{Harsoula} \& {Kalapotharakos}}{{Harsoula} \&
  {Kalapotharakos}}{2009}]{2009MNRAS.394.1605H}
{Harsoula} M.,  {Kalapotharakos} C.,  2009, \mn@doi [\mnras]
  {10.1111/j.1365-2966.2009.14427.x}, \href
  {https://ui.adsabs.harvard.edu/abs/2009MNRAS.394.1605H} {394, 1605}

\bibitem[\protect\citeauthoryear{{Hernquist} \& {Ostriker}}{{Hernquist} \&
  {Ostriker}}{1992}]{1992ApJ...386..375H}
{Hernquist} L.,  {Ostriker} J.~P.,  1992, \mn@doi [\apj] {10.1086/171025},
  \href {https://ui.adsabs.harvard.edu/abs/1992ApJ...386..375H} {386, 375}

\bibitem[\protect\citeauthoryear{{Kaufmann} \& {Contopoulos}}{{Kaufmann} \&
  {Contopoulos}}{1996}]{1996A&A...309..381K}
{Kaufmann} D.~E.,  {Contopoulos} G.,  1996, \aap, \href
  {https://ui.adsabs.harvard.edu/abs/1996A&A...309..381K} {309, 381}

\bibitem[\protect\citeauthoryear{{Kuijken} \& {Dubinski}}{{Kuijken} \&
  {Dubinski}}{1995}]{1995MNRAS.277.1341K}
{Kuijken} K.,  {Dubinski} J.,  1995, \mn@doi [\mnras]
  {10.1093/mnras/277.4.1341}, \href
  {https://ui.adsabs.harvard.edu/abs/1995MNRAS.277.1341K} {277, 1341}

\bibitem[\protect\citeauthoryear{{Laskar}}{{Laskar}}{1990}]{1990Icar...88..266L}
{Laskar} J.,  1990, \mn@doi [\icarus] {10.1016/0019-1035(90)90084-M}, \href
  {https://ui.adsabs.harvard.edu/abs/1990Icar...88..266L} {88, 266}

\bibitem[\protect\citeauthoryear{{Laurikainen} \& {Salo}}{{Laurikainen} \&
  {Salo}}{2017}]{2017A&A...598A..10L}
{Laurikainen} E.,  {Salo} H.,  2017, \mn@doi [\aap]
  {10.1051/0004-6361/201628936}, \href
  {https://ui.adsabs.harvard.edu/abs/2017A&A...598A..10L} {598, A10}

\bibitem[\protect\citeauthoryear{{Li}, {Ho}  \& {Barth}}{{Li}
  et~al.}{2017}]{2017ApJ...845...87L}
{Li} Z.-Y.,  {Ho} L.~C.,   {Barth} A.~J.,  2017, \mn@doi [\apj]
  {10.3847/1538-4357/aa7fba}, \href
  {https://ui.adsabs.harvard.edu/abs/2017ApJ...845...87L} {845, 87}

\bibitem[\protect\citeauthoryear{{Lilley}, {Sanders}, {Evans}  \&
  {Erkal}}{{Lilley} et~al.}{2018}]{2018MNRAS.476.2092L}
{Lilley} E.~J.,  {Sanders} J.~L.,  {Evans} N.~W.,   {Erkal} D.,  2018, \mn@doi
  [\mnras] {10.1093/mnras/sty296}, \href
  {https://ui.adsabs.harvard.edu/abs/2018MNRAS.476.2092L} {476, 2092}

\bibitem[\protect\citeauthoryear{{{\L}okas}}{{{\L}okas}}{2019}]{2019A&A...629A..52L}
{{\L}okas} E.~L.,  2019, \mn@doi [\aap] {10.1051/0004-6361/201936056}, \href
  {https://ui.adsabs.harvard.edu/abs/2019A&A...629A..52L} {629, A52}

\bibitem[\protect\citeauthoryear{{L{\"u}tticke}, {Dettmar}  \&
  {Pohlen}}{{L{\"u}tticke} et~al.}{2000}]{2000A&AS..145..405L}
{L{\"u}tticke} R.,  {Dettmar} R.~J.,   {Pohlen} M.,  2000, \mn@doi [\aaps]
  {10.1051/aas:2000354}, \href
  {https://ui.adsabs.harvard.edu/abs/2000A&AS..145..405L} {145, 405}

\bibitem[\protect\citeauthoryear{Lynden-Bell}{Lynden-Bell}{1996}]{1996LNP...474....7L}
Lynden-Bell D.,  1996, in , Barred galaxies and circumnuclear activity.
Springer, pp 7--18

\bibitem[\protect\citeauthoryear{{Marinova} \& {Jogee}}{{Marinova} \&
  {Jogee}}{2007}]{2007ApJ...659.1176M}
{Marinova} I.,  {Jogee} S.,  2007, \mn@doi [\apj] {10.1086/512355}, \href
  {http://adsabs.harvard.edu/abs/2007ApJ...659.1176M} {659, 1176}

\bibitem[\protect\citeauthoryear{{Martinez-Valpuesta} \&
  {Shlosman}}{{Martinez-Valpuesta} \& {Shlosman}}{2004}]{2004ApJ...613L..29M}
{Martinez-Valpuesta} I.,  {Shlosman} I.,  2004, \mn@doi [\apjl]
  {10.1086/424876}, \href
  {https://ui.adsabs.harvard.edu/abs/2004ApJ...613L..29M} {613, L29}

\bibitem[\protect\citeauthoryear{{Martinez-Valpuesta}, {Shlosman}  \&
  {Heller}}{{Martinez-Valpuesta} et~al.}{2006}]{2006ApJ...637..214M}
{Martinez-Valpuesta} I.,  {Shlosman} I.,   {Heller} C.,  2006, \mn@doi [\apj]
  {10.1086/498338}, \href
  {https://ui.adsabs.harvard.edu/abs/2006ApJ...637..214M} {637, 214}

\bibitem[\protect\citeauthoryear{{Merritt} \& {Sellwood}}{{Merritt} \&
  {Sellwood}}{1994}]{1994ApJ...425..551M}
{Merritt} D.,  {Sellwood} J.~A.,  1994, \mn@doi [\apj] {10.1086/174005}, \href
  {https://ui.adsabs.harvard.edu/abs/1994ApJ...425..551M} {425, 551}

\bibitem[\protect\citeauthoryear{{Muzzio}, {Carpintero}  \& {Wachlin}}{{Muzzio}
  et~al.}{2005}]{2005CeMDA..91..173M}
{Muzzio} J.~C.,  {Carpintero} D.~D.,   {Wachlin} F.~C.,  2005, \mn@doi
  [Celestial Mechanics and Dynamical Astronomy] {10.1007/s10569-005-1608-4},
  \href {https://ui.adsabs.harvard.edu/abs/2005CeMDA..91..173M} {91, 173}

\bibitem[\protect\citeauthoryear{{Navarro}, {Frenk}  \& {White}}{{Navarro}
  et~al.}{1996}]{1996ApJ...462..563N}
{Navarro} J.~F.,  {Frenk} C.~S.,   {White} S. D.~M.,  1996, \mn@doi [\apj]
  {10.1086/177173}, \href
  {https://ui.adsabs.harvard.edu/abs/1996ApJ...462..563N} {462, 563}

\bibitem[\protect\citeauthoryear{{Navarro}, {Frenk}  \& {White}}{{Navarro}
  et~al.}{1997}]{1997ApJ...490..493N}
{Navarro} J.~F.,  {Frenk} C.~S.,   {White} S. D.~M.,  1997, \mn@doi [\apj]
  {10.1086/304888}, \href
  {https://ui.adsabs.harvard.edu/abs/1997ApJ...490..493N} {490, 493}

\bibitem[\protect\citeauthoryear{Oppenheim, Willsky, Nawab  \& Ding}{Oppenheim
  et~al.}{1997}]{oppenheim1997signals}
Oppenheim A.~V.,  Willsky A.~S.,  Nawab S.~H.,   Ding J.-J.,  1997, Signals and
  systems.
 Second Edition Vol. 2, Prentice hall Upper Saddle River, NJ

\bibitem[\protect\citeauthoryear{{Parul}, {Smirnov}  \& {Sotnikova}}{{Parul}
  et~al.}{2020}]{2020ApJ...895...12P}
{Parul} H.~D.,  {Smirnov} A.~A.,   {Sotnikova} N.~Y.,  2020, \mn@doi [\apj]
  {10.3847/1538-4357/ab76ce}, \href
  {https://ui.adsabs.harvard.edu/abs/2020ApJ...895...12P} {895, 12}

\bibitem[\protect\citeauthoryear{{Patsis} \& {Athanassoula}}{{Patsis} \&
  {Athanassoula}}{2019}]{2019MNRAS.490.2740P}
{Patsis} P.~A.,  {Athanassoula} E.,  2019, \mn@doi [\mnras]
  {10.1093/mnras/stz2588}, \href
  {https://ui.adsabs.harvard.edu/abs/2019MNRAS.490.2740P} {490, 2740}

\bibitem[\protect\citeauthoryear{{Patsis} \& {Harsoula}}{{Patsis} \&
  {Harsoula}}{2018}]{2018A&A...612A.114P}
{Patsis} P.~A.,  {Harsoula} M.,  2018, \mn@doi [\aap]
  {10.1051/0004-6361/201731114}, \href
  {https://ui.adsabs.harvard.edu/abs/2018A&A...612A.114P} {612, A114}

\bibitem[\protect\citeauthoryear{{Patsis} \& {Katsanikas}}{{Patsis} \&
  {Katsanikas}}{2014}]{2014MNRAS.445.3546P}
{Patsis} P.~A.,  {Katsanikas} M.,  2014, \mn@doi [\mnras]
  {10.1093/mnras/stu1970}, \href
  {https://ui.adsabs.harvard.edu/abs/2014MNRAS.445.3546P} {445, 3546}

\bibitem[\protect\citeauthoryear{{Patsis}, {Athanassoula}  \&
  {Quillen}}{{Patsis} et~al.}{1997}]{1997ApJ...483..731P}
{Patsis} P.~A.,  {Athanassoula} E.,   {Quillen} A.~C.,  1997, \mn@doi [\apj]
  {10.1086/304287}, \href
  {https://ui.adsabs.harvard.edu/abs/1997ApJ...483..731P} {483, 731}

\bibitem[\protect\citeauthoryear{{Patsis}, {Skokos}  \&
  {Athanassoula}}{{Patsis} et~al.}{2002}]{2002MNRAS.337..578P}
{Patsis} P.~A.,  {Skokos} C.,   {Athanassoula} E.,  2002, \mn@doi [\mnras]
  {10.1046/j.1365-8711.2002.05943.x}, \href
  {https://ui.adsabs.harvard.edu/abs/2002MNRAS.337..578P} {337, 578}

\bibitem[\protect\citeauthoryear{{Patsis}, {Kalapotharakos}  \&
  {Grosb{\o}l}}{{Patsis} et~al.}{2010}]{2010MNRAS.408...22P}
{Patsis} P.~A.,  {Kalapotharakos} C.,   {Grosb{\o}l} P.,  2010, \mn@doi
  [\mnras] {10.1111/j.1365-2966.2010.17062.x}, \href
  {https://ui.adsabs.harvard.edu/abs/2010MNRAS.408...22P} {408, 22}

\bibitem[\protect\citeauthoryear{{Polyachenko} \& {Polyachenko}}{{Polyachenko}
  \& {Polyachenko}}{2003}]{2003AstL...29..447P}
{Polyachenko} V.~L.,  {Polyachenko} E.~V.,  2003, \mn@doi [Astronomy Letters]
  {10.1134/1.1589862}, \href
  {http://adsabs.harvard.edu/abs/2003AstL...29..447P} {29, 447}

\bibitem[\protect\citeauthoryear{{Portail}, {Wegg}  \& {Gerhard}}{{Portail}
  et~al.}{2015}]{2015MNRAS.450L..66P}
{Portail} M.,  {Wegg} C.,   {Gerhard} O.,  2015, \mn@doi [\mnras]
  {10.1093/mnrasl/slv048}, \href
  {https://ui.adsabs.harvard.edu/abs/2015MNRAS.450L..66P} {450, L66}

\bibitem[\protect\citeauthoryear{Press, Teukolsky, Flannery  \&
  Vetterling}{Press et~al.}{1992}]{press1992numerical}
Press W.~H.,  Teukolsky S.~A.,  Flannery B.~P.,   Vetterling W.~T.,  1992,
  Numerical recipes in Fortran 77: volume 1, volume 1 of Fortran numerical
  recipes: the art of scientific computing.
Cambridge university press

\bibitem[\protect\citeauthoryear{{Raha}, {Sellwood}, {James}  \& {Kahn}}{{Raha}
  et~al.}{1991}]{1991Natur.352..411R}
{Raha} N.,  {Sellwood} J.~A.,  {James} R.~A.,   {Kahn} F.~D.,  1991, \mn@doi
  [\nat] {10.1038/352411a0}, \href
  {https://ui.adsabs.harvard.edu/abs/1991Natur.352..411R} {352, 411}

\bibitem[\protect\citeauthoryear{Richmond-Navarro, Barquero-Mena,
  Sol{\'\i}s-Villalta  \& Palma-Quir{\'o}s}{Richmond-Navarro
  et~al.}{2017}]{richmond2017interpolacion}
Richmond-Navarro G.,  Barquero-Mena T.~G.,  Sol{\'\i}s-Villalta O.~M.,
  Palma-Quir{\'o}s D.~M.,  2017, Revista Tecnolog{\'\i}a en Marcha, 30, 14

\bibitem[\protect\citeauthoryear{{Saha}, {Pfenniger}  \& {Taam}}{{Saha}
  et~al.}{2013}]{2013ApJ...764..123S}
{Saha} K.,  {Pfenniger} D.,   {Taam} R.~E.,  2013, \mn@doi [\apj]
  {10.1088/0004-637X/764/2/123}, \href
  {https://ui.adsabs.harvard.edu/abs/2013ApJ...764..123S} {764, 123}

\bibitem[\protect\citeauthoryear{{Skokos}, {Patsis}  \&
  {Athanassoula}}{{Skokos} et~al.}{2002a}]{2002MNRAS.333..847S}
{Skokos} C.,  {Patsis} P.~A.,   {Athanassoula} E.,  2002a, \mn@doi [\mnras]
  {10.1046/j.1365-8711.2002.05468.x}, \href
  {https://ui.adsabs.harvard.edu/abs/2002MNRAS.333..847S} {333, 847}

\bibitem[\protect\citeauthoryear{{Skokos}, {Patsis}  \&
  {Athanassoula}}{{Skokos} et~al.}{2002b}]{2002MNRAS.333..861S}
{Skokos} C.,  {Patsis} P.~A.,   {Athanassoula} E.,  2002b, \mn@doi [\mnras]
  {10.1046/j.1365-8711.2002.05469.x}, \href
  {https://ui.adsabs.harvard.edu/abs/2002MNRAS.333..861S} {333, 861}

\bibitem[\protect\citeauthoryear{{Smirnov} \& {Sotnikova}}{{Smirnov} \&
  {Sotnikova}}{2018}]{2018MNRAS.481.4058S}
{Smirnov} A.~A.,  {Sotnikova} N.~Y.,  2018, \mn@doi [\mnras]
  {10.1093/mnras/sty2423}, \href
  {https://ui.adsabs.harvard.edu/abs/2018MNRAS.481.4058S} {481, 4058}

\bibitem[\protect\citeauthoryear{{Smirnov} \& {Sotnikova}}{{Smirnov} \&
  {Sotnikova}}{2019}]{2019MNRAS.485.1900S}
{Smirnov} A.~A.,  {Sotnikova} N.~Y.,  2019, \mn@doi [\mnras]
  {10.1093/mnras/stz546}, \href
  {https://ui.adsabs.harvard.edu/abs/2019MNRAS.485.1900S} {485, 1900}

\bibitem[\protect\citeauthoryear{{Smirnov}, {Tikhonenko}  \&
  {Sotnikova}}{{Smirnov} et~al.}{2021}]{2021MNRAS.502.4689S}
{Smirnov} A.~A.,  {Tikhonenko} I.~S.,   {Sotnikova} N.~Y.,  2021, \mn@doi
  [\mnras] {10.1093/mnras/stab327}, \href
  {https://ui.adsabs.harvard.edu/abs/2021MNRAS.502.4689S} {502, 4689}

\bibitem[\protect\citeauthoryear{{Springel}}{{Springel}}{2005}]{2005MNRAS.364.1105S}
{Springel} V.,  2005, \mn@doi [\mnras] {10.1111/j.1365-2966.2005.09655.x},
  \href {https://ui.adsabs.harvard.edu/abs/2005MNRAS.364.1105S} {364, 1105}

\bibitem[\protect\citeauthoryear{{Springel} \& {White}}{{Springel} \&
  {White}}{1999}]{1999MNRAS.307..162S}
{Springel} V.,  {White} S. D.~M.,  1999, \mn@doi [\mnras]
  {10.1046/j.1365-8711.1999.02613.x}, \href
  {https://ui.adsabs.harvard.edu/abs/1999MNRAS.307..162S} {307, 162}

\bibitem[\protect\citeauthoryear{{Springel}, {Yoshida}  \& {White}}{{Springel}
  et~al.}{2001}]{2001NewA....6...79S}
{Springel} V.,  {Yoshida} N.,   {White} S. D.~M.,  2001, \mn@doi [\na]
  {10.1016/S1384-1076(01)00042-2}, \href
  {https://ui.adsabs.harvard.edu/abs/2001NewA....6...79S} {6, 79}

\bibitem[\protect\citeauthoryear{{Tsigaridi} \& {Patsis}}{{Tsigaridi} \&
  {Patsis}}{2015}]{2015MNRAS.448.3081T}
{Tsigaridi} L.,  {Patsis} P.~A.,  2015, \mn@doi [\mnras]
  {10.1093/mnras/stv206}, \href
  {https://ui.adsabs.harvard.edu/abs/2015MNRAS.448.3081T} {448, 3081}

\bibitem[\protect\citeauthoryear{{Valencia-Enr{\'\i}quez}, {Puerari}  \&
  {Rodrigues}}{{Valencia-Enr{\'\i}quez} et~al.}{2019}]{2019AJ....157..175V}
{Valencia-Enr{\'\i}quez} D.,  {Puerari} I.,   {Rodrigues} I.,  2019, \mn@doi
  [\aj] {10.3847/1538-3881/ab100f}, \href
  {https://ui.adsabs.harvard.edu/abs/2019AJ....157..175V} {157, 175}

\bibitem[\protect\citeauthoryear{{Valluri}, {Shen}, {Abbott}  \&
  {Debattista}}{{Valluri} et~al.}{2016}]{2016ApJ...818..141V}
{Valluri} M.,  {Shen} J.,  {Abbott} C.,   {Debattista} V.~P.,  2016, \mn@doi
  [\apj] {10.3847/0004-637X/818/2/141}, \href
  {https://ui.adsabs.harvard.edu/abs/2016ApJ...818..141V} {818, 141}

\bibitem[\protect\citeauthoryear{{Vasiliev}}{{Vasiliev}}{2019}]{2019MNRAS.482.1525V}
{Vasiliev} E.,  2019, \mn@doi [\mnras] {10.1093/mnras/sty2672}, \href
  {https://ui.adsabs.harvard.edu/abs/2019MNRAS.482.1525V} {482, 1525}

\bibitem[\protect\citeauthoryear{{Vasiliev} \& {Athanassoula}}{{Vasiliev} \&
  {Athanassoula}}{2015}]{2015MNRAS.450.2842V}
{Vasiliev} E.,  {Athanassoula} E.,  2015, \mn@doi [\mnras]
  {10.1093/mnras/stv805}, \href
  {https://ui.adsabs.harvard.edu/abs/2015MNRAS.450.2842V} {450, 2842}

\bibitem[\protect\citeauthoryear{{Voglis}, {Harsoula}  \&
  {Contopoulos}}{{Voglis} et~al.}{2007}]{2007MNRAS.381..757V}
{Voglis} N.,  {Harsoula} M.,   {Contopoulos} G.,  2007, \mn@doi [\mnras]
  {10.1111/j.1365-2966.2007.12263.x}, \href
  {https://ui.adsabs.harvard.edu/abs/2007MNRAS.381..757V} {381, 757}

\bibitem[\protect\citeauthoryear{{Wang}, {Athanassoula}  \& {Mao}}{{Wang}
  et~al.}{2016}]{2016MNRAS.463.3499W}
{Wang} Y.,  {Athanassoula} E.,   {Mao} S.,  2016, \mn@doi [\mnras]
  {10.1093/mnras/stw2301}, \href
  {https://ui.adsabs.harvard.edu/abs/2016MNRAS.463.3499W} {463, 3499}

\bibitem[\protect\citeauthoryear{{Wang}, {Athanassoula}  \& {Mao}}{{Wang}
  et~al.}{2020}]{2020A&A...639A..38W}
{Wang} Y.,  {Athanassoula} E.,   {Mao} S.,  2020, \mn@doi [\aap]
  {10.1051/0004-6361/202038225}, \href
  {https://ui.adsabs.harvard.edu/abs/2020A&A...639A..38W} {639, A38}

\bibitem[\protect\citeauthoryear{{Wang}, {Athanassoula}, {Patsis}  \&
  {Mao}}{{Wang} et~al.}{2022}]{2022A&A...668A..55W}
{Wang} Y.,  {Athanassoula} E.,  {Patsis} P.,   {Mao} S.,  2022, \mn@doi [\aap]
  {10.1051/0004-6361/202243699}, \href
  {https://ui.adsabs.harvard.edu/abs/2022A&A...668A..55W} {668, A55}

\bibitem[\protect\citeauthoryear{{Wozniak}}{{Wozniak}}{1994}]{1994LNP...430..264W}
{Wozniak} H.,  1994, {Regular Orbits and Cantori in the Potential of the Barred
  Galaxy NGC 936}.
Springer Berlin Heidelberg, p.~264, \mn@doi{10.1007/BFb0058117}

\bibitem[\protect\citeauthoryear{{Wozniak} \& {Pfenniger}}{{Wozniak} \&
  {Pfenniger}}{1999}]{1999CeMDA..73..149W}
{Wozniak} H.,  {Pfenniger} D.,  1999, \mn@doi [Celestial Mechanics and
  Dynamical Astronomy] {10.1023/A:1008394929716}, \href
  {https://ui.adsabs.harvard.edu/abs/1999CeMDA..73..149W} {73, 149}

\makeatother
\end{thebibliography}




\appendix
\section{Frozen potential and integration}

The methodology employed to determine the 3D 'frozen' potential $V(x,y,z)$ was based on creating a three-dimensional cubic grid that encompasses the model. At each grid point, we calculated the potential, which results from all the particles present. We obtained the potential at the specific position of the particle being integrated to determine its orbital trajectory. This potential was obtained through an interpolation process, while the force components were derived from this interpolation. In our analysis, we operate within the co-rotating frame of reference aligned with the bar. Utilizing a frozen potential, our model becomes an independent Hamiltonian system, and we can formulate its Hamiltonian as follows:
\begin{equation}
    H=\frac{1}{2}(\dot{x}^2+\dot{y}^2+\dot{z}^2)+V(x,y,z)-\frac{1}{2}\Omega_p^2(x^2+y^2)
\end{equation}

where $(x, y, z)$ denotes the coordinates in a Cartesian frame of reference that rotates clockwise around the z-axis with an angular velocity of $\Omega_p$. Finally, the temporal evolution was determined using the fourth order Runge-Kutta method.

The proposed approach for three-dimensional interpolation involves the utilization of a cubic three-dimensional grid \citep{richmond2017interpolacion}, see equation \ref{eq:intp}. This entails that every point in the grid is surrounded by eight neighboring points positioned at the vertices. Each vertex represents a known value, thereby enabling the interpolation process. The method considers the proximity to the known points, implying that a point located closer to the point of interest will exert a more significant influence on the resulting interpolation outcome.

\begin{equation}
    \begin{split}
      A(x,y,z) = &   \left(\frac{x_1-x}{x_1-x_0}\right)
                     \left(\frac{y_1-y}{y_1-y_0}\right)
                     \left(\frac{z_1-z}{z_1-z_0}\right)A(x_0,y_0,z_0)\\          
                 & + \left(\frac{x_1-x}{x_1-x_0}\right)
                     \left(\frac{y_1-y}{y_1-y_0}\right)
                     \left(\frac{z-z_0}{z_1-z_0}\right)A(x_0,y_0,z_1)\\
                 & + \left(\frac{x_1-x}{x_1-x_0}\right)
                     \left(\frac{y-y_0}{y_1-y_0}\right)
                     \left(\frac{z_1-z}{z_1-z_0}\right)A(x_0,y_1,z_0)\\
                 & + \left(\frac{x_1-x}{x_1-x_0}\right)
                     \left(\frac{y-y_0}{y_1-y_0}\right)
                     \left(\frac{z-z_0}{z_1-z_0}\right)A(x_0,y_1,z_1)\\
                &  + \left(\frac{x-x_0}{x_1-x_0}\right)
                     \left(\frac{y_1-y}{y_1-y_0}\right)
                     \left(\frac{z_1-z}{z_1-z_0}\right)A(x_1,y_0,z_0)\\          
                 & + \left(\frac{x-x_0}{x_1-x_0}\right)
                     \left(\frac{y_1-y}{y_1-y_0}\right)
                     \left(\frac{z-z_0}{z_1-z_0}\right)A(x_1,y_0,z_1)\\
                 & + \left(\frac{x-x_0}{x_1-x_0}\right)
                     \left(\frac{y-y_0}{y_1-y_0}\right)
                     \left(\frac{z_1-z}{z_1-z_0}\right)A(x_1,y_1,z_0)\\
                 & + \left(\frac{x-x_0}{x_1-x_0}\right)
                     \left(\frac{y-y_0}{y_1-y_0}\right)
                     \left(\frac{z-z_0}{z_1-z_0}\right)A(x_1,y_1,z_1)
    \end{split}
    \label{eq:intp}
\end{equation}

The initial conditions for the calculation of the orbits were taken from the positions and velocities of the N-body disk particles at the same snapshot of the calculated potential.

\label{appendix0}

\section{Individual density maps}
\label{appendixa}

In this appendix, we present the density maps (see Figure \ref{fig:maps}), but now for each bin on $\omega_z/\omega_x$ as indicated in Figure \ref{fig:frozenL03}, \ref{fig:frozenL04}, and \ref{fig:frozenL05}.

\begin{figure*}
	\includegraphics[scale=0.8]{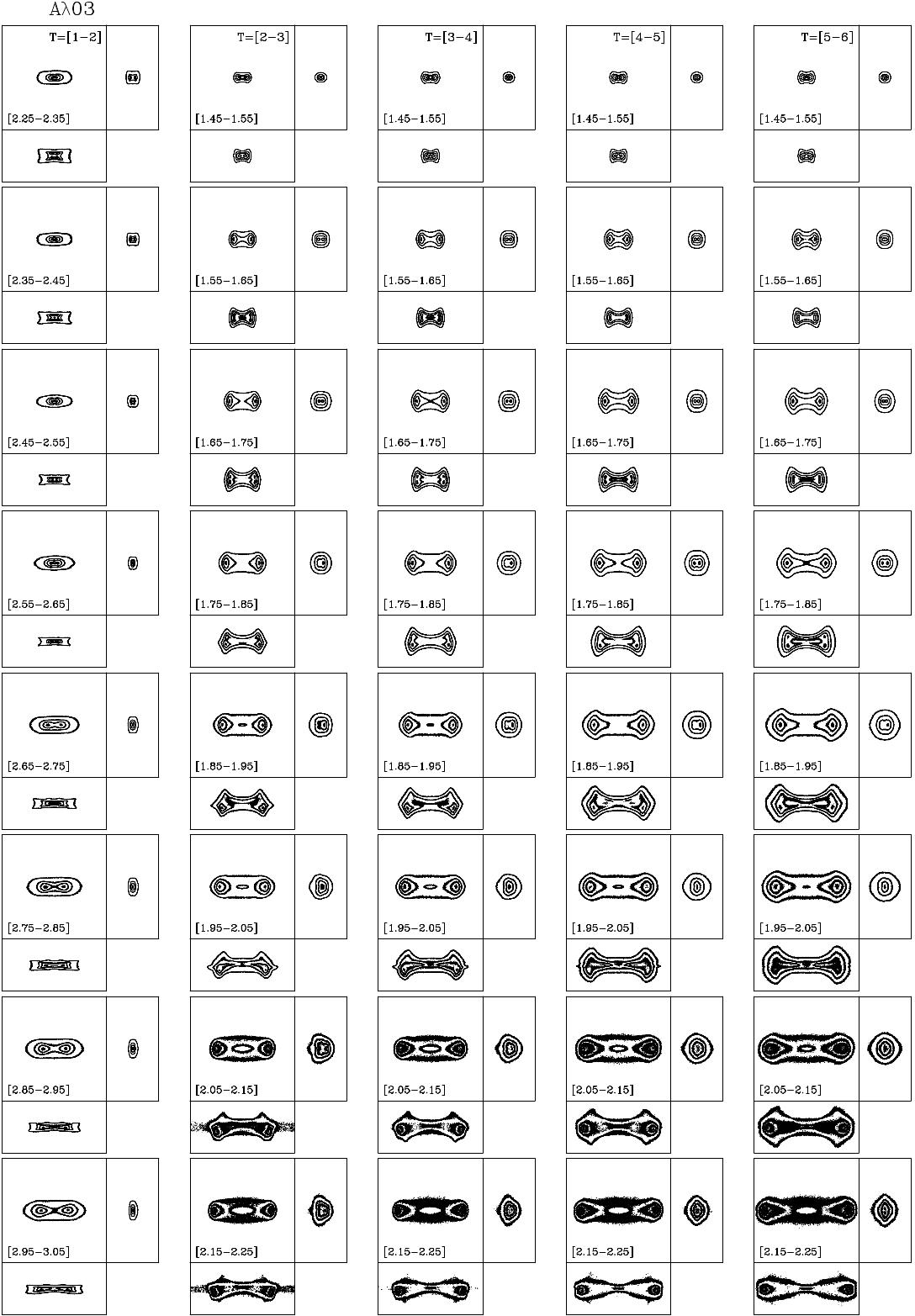}
	\caption{Orbits assembly density profile (OADP) in face-on, edge-on, and end-on views for disk particles of the $A\lambda03$ model having $1.9<\omega_R/\omega_x<2.1$ and $\omega_z/\omega_x$ as indicated in each panel. The time increases from left to right in units of one Gyr; the x and y axes are from -5.0 to 5.0 kpc, while z is from -2.5 to 2.5 kpc. The eight bins of the ratio $\omega_z/\omega_x$ shown here are those selected in Figure \ref{fig:frozenL03}. For this model, the bins from time interval T=[2-3] to T=[5-6] are at the same frequencies, so we can see the time evolution of the structure created by that particular $\omega_z/\omega_x$ bin (see text).}
	\label{fig:L03_map}
\end{figure*}

Figure \ref{fig:L03_map} shows orbits assembly density profile (OADP) maps in their different views for B-type orbits from the $A\lambda03$ model. The frequency ratio interval ($\omega_z/\omega_x$) shows on the bottom left of each panel. We note that before the buckling phase, all the B-type orbits in $\omega_z/\omega_x>2.25$ contribute to the shape of the bar structure in its initial state, which is the boxy shape (the first column, T=[1-2]). When the bar buckles, these types of orbits no longer contribute to the form of the bar. The orbits evolve and accommodate in the frequency interval $1.45\leq\omega_z/\omega_x\leq2.25$ (T=[2-3] to T=[5-6]), but the largest contributions to the bar structure in this model come from orbits in the lower part of this interval (see Figures \ref{fig:frozenL03}). These orbits with low $\omega_z/\omega_x$ are located in the internal part of the model, where the density of the model is high. They evolve to a small and symmetric peanut shape in the face-on and edge-on views. These orbit families all evolve in a similar way.
The evolution of orbits with larger $\omega_z/\omega_x$ are a bit different. The size of these stellar orbits increases, and they are responsible for the formation of the buckle. But in this case, firstly, the orbits bend up showing a ``down smile'' figure. Later on, these orbits evolve to a more symmetric figure, and the buckle becomes symmetric with respect to the disk plane.

\begin{figure*}
	\includegraphics[scale=0.8]{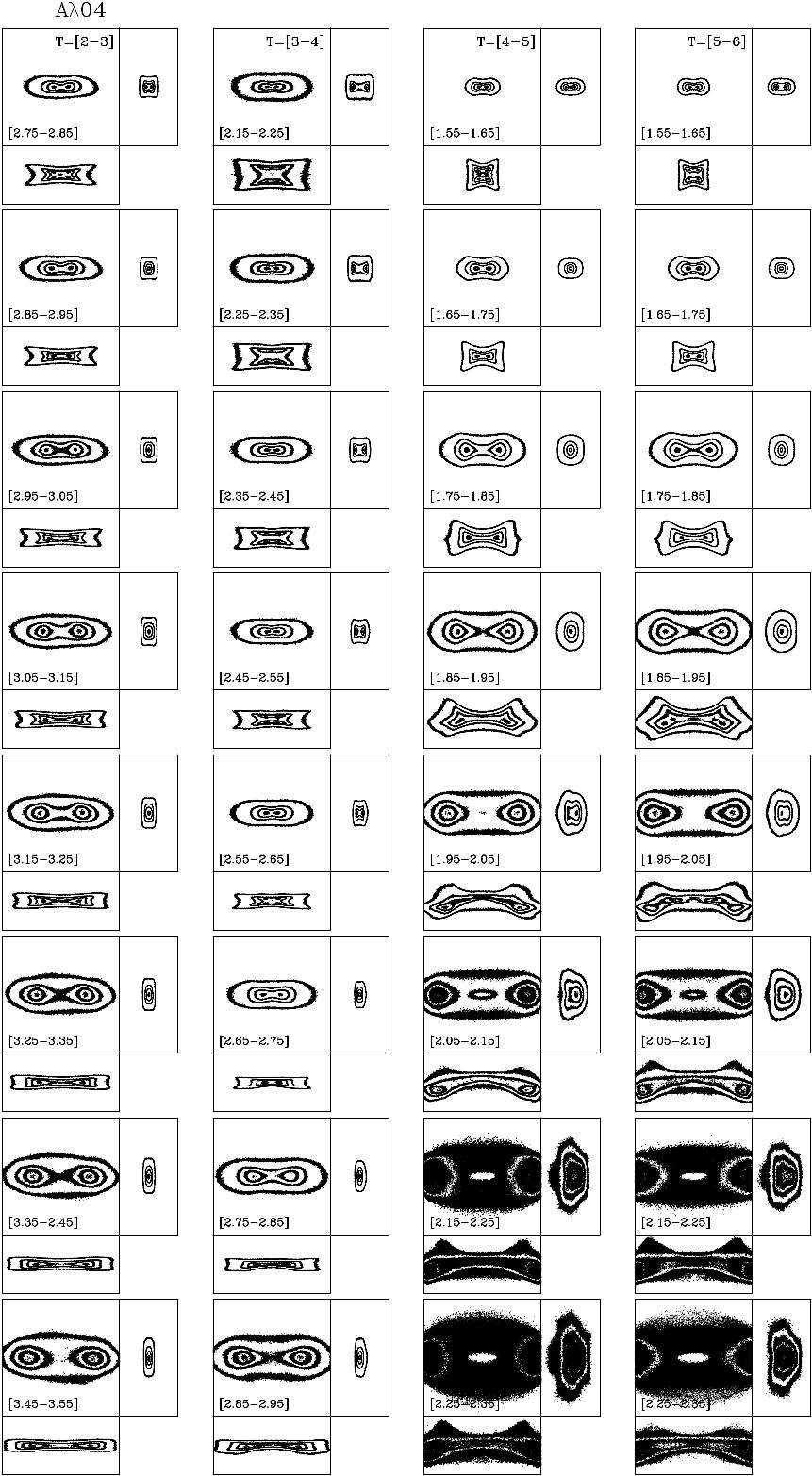}
	\caption{As in Figure \ref{fig:L03_map}, but for the $A\lambda04$ model (see text).}
	\label{fig:L04_map}
\end{figure*}

We present the OADP for the $A\lambda04$ model in Figure \ref{fig:L04_map}. This figure shows the OADP from 2 to 6 Gyrs. For this model, the B-type orbits that begin to contribute to the bar structure appear in the interval $\omega_z/\omega_x>2.75$ (the first column, T=[2-3]). Later on, the bar evolves and the buckle begins to appear. Similar to the $A\lambda03$ model, the orbits with a frequency ratio larger than 1.85 participate in the buckle structure. But here, as the bar formation was delayed compared to the $A\lambda03$ model, the buckle has not achieved its final configuration at the time of the simulation.

\begin{figure*}
	\includegraphics[scale=0.8]{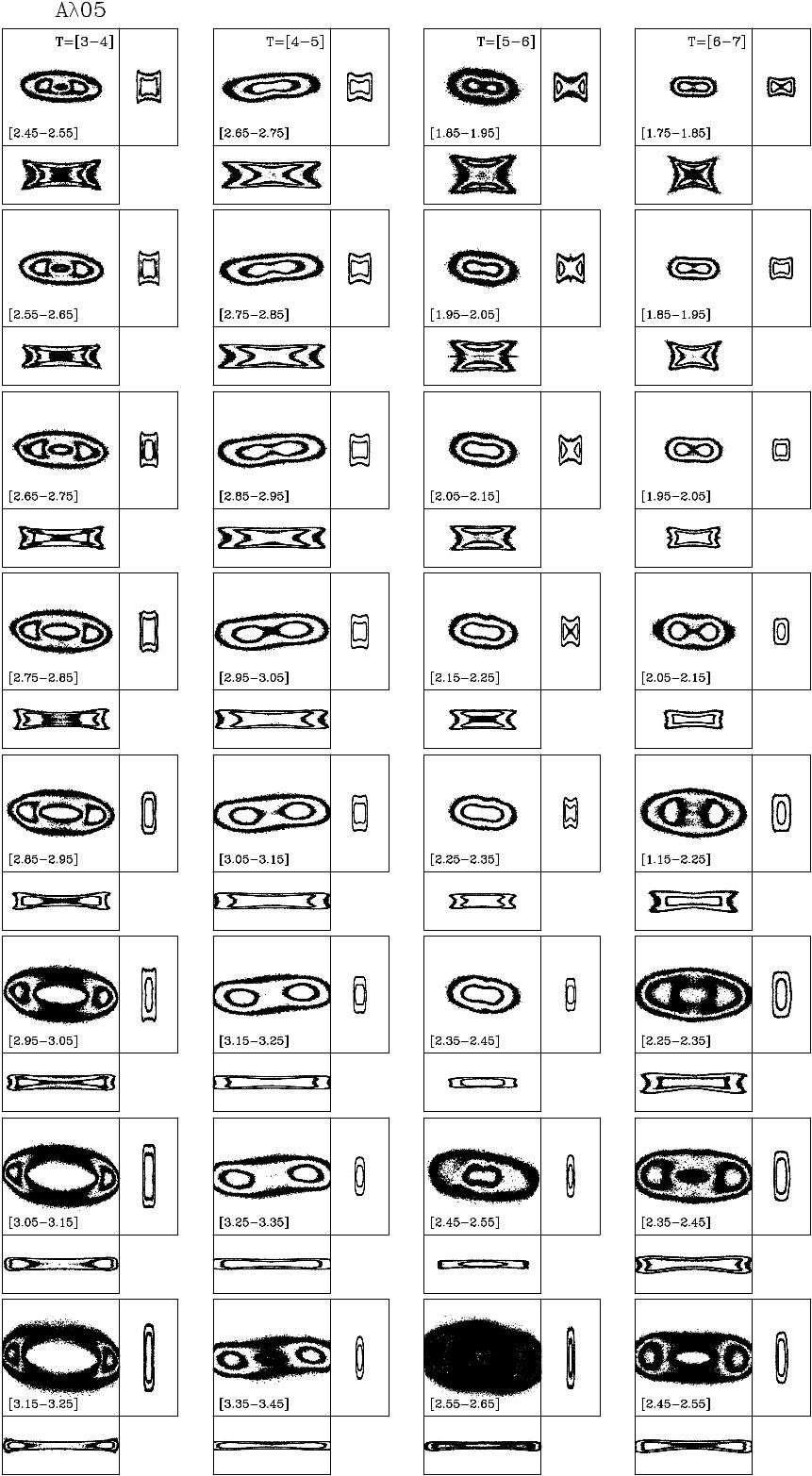}
	\caption{As in Figure \ref{fig:L03_map}, but for the $A\lambda05$ model (see text).}
	\label{fig:L05_map}
\end{figure*}

Finally, we present the OADP for the $A\lambda05$ model, depicted in Figure \ref{fig:L05_map}. In this model, significantly fewer particles exhibit $1.9<\omega_R/\omega_x<2.1$ (refer to Figure \ref{fig:frozenL05}). This halo-dominated model lacks peanut shapes in the OADP maps; instead, it displays boxy and X-shaped patterns in OADP edge-on views. Moreover, due to the dominant contribution of the halo, the model does not reach the buckling phase during our simulation. As we note in the former models, stellar orbits need to have a frequency ratio $\omega_z/\omega_x$ around 2.0 for the buckling appearance. Although some orbits show this frequency ratio, they take longer to buckle out of the disk plane.

\bsp	
\label{lastpage}
\end{document}